\documentclass[english,aps,superscriptaddress,prb,eqsecnum,notitlepage]{revtex4-1}
\usepackage{babel}
\usepackage{amsmath,amssymb}
\usepackage{leftidx}
\usepackage{graphicx}
\usepackage{color}
\usepackage{xcolor}
\usepackage{epsfig}
\usepackage{ulem}
\usepackage{youngtab}

\newcommand{\bea}{\begin{eqnarray}}
\newcommand{\eea}{\end{eqnarray}}

\newcommand{\perm}[1]{\mathcal{P}_{#1}}

\newcommand{\tr}{\mbox{tr}}
\newcommand{\vphi}{\boldsymbol{\phi}}
\newcommand{\vvphi}{\boldsymbol{\varphi}}
\newcommand{\bDelta}{\boldsymbol{\Delta}}

\newcommand{\fO}{\mathcal{O}}
\usepackage{color}
\usepackage{tikz-feynman}
\newcommand{\be}{\begin{equation}}
\newcommand{\ee}{\end{equation}}

\newcommand{\fL}{\mathcal{L}}
\newcommand{\fD}{\mathcal{D}}
\newcommand{\fS}{\mathcal{S}}

\begin{document}
\title{Generalization of the Haldane conjecture to SU($n$) chains}
\date{\today}
\author{Kyle Wamer}
\affiliation{Department of Physics and Astronomy and Stewart Blusson Quantum Matter Institute, University of British Columbia, 
Vancouver, B.C., Canada, V6T1Z1}
\author{Mikl\'os Lajk\'o}
\affiliation{Institute of Physics, Ecole Polytechnique F\'ed\'erale de Lausanne (EPFL), CH-1015, Lausanne, Switzerland}
\author{Fr\'ed\'eric Mila}
\affiliation{Institute of Physics, Ecole Polytechnique F\'ed\'erale de Lausanne (EPFL), CH-1015, Lausanne, Switzerland}
\author{Ian Affleck}
\affiliation{Department of Physics and Astronomy and Stewart Blusson Quantum Matter Institute, University of British Columbia, 
Vancouver, B.C., Canada, V6T1Z1}

\begin{abstract}

Recently, SU(3) chains in the symmetric and self-conjugate representations have been studied using field theory techniques. For certain representations, namely rank-$p$ symmetric ones with $p$ not a multiple of 3, it was argued that the ground state exhibits gapless excitations. For the remaining representations considered, a finite energy gap exists above the ground state. In this paper, we extend these results to SU($n$) chains in the symmetric representation. For a rank-$p$ symmetric representation with $n$ and $p$ coprime, we predict gapless excitations above the ground state. If $p$ is a multiple of $n$, we predict a unique ground state with a finite energy gap. Finally, if $p$ and $n$ have a greatest common divisor $1<q<n$, we predict a ground state degeneracy of $n/q$, with a finite energy gap. To arrive at these results, we derive a non-Lorentz invariant flag manifold sigma model description of the SU($n$) chains, and use the renormalization group to show that Lorentz invariance is restored at low energies. We then make use of recently developed anomaly matching conditions for these Lorentz-invariant models. We also review the Lieb-Shultz-Mattis-Affleck theorem, and extend it to SU($n$) models with longer range interactions.
\end{abstract}
\maketitle
\section{Introduction}

In 1983, Haldane showed that in the limit of large spin, the antiferromagnetic spin chain maps to a relativistic field theory with topological term proportional to $2\pi s$.\cite{HaldanePRL1983, HaldanePLA1983} He then argued that integer spin chains are gapped and have exponentially decaying correlation functions, while half-integer spin chains are gapless, with power-law correlations. This became known as ``Haldane's conjecture''. While these arguments followed from a large spin limit, this conjecture has been verified experimentally for quasi-one dimensional $s=1$ chains\cite{PhysRevLett.56.371, doi:10.1002/9783527620548.ch2}, and numerically for spins up to $s=4$.\cite{PhysRevB.28.3914, PhysRevB.33.659, Kennedy_1990, PhysRevB.48.3844, PhysRevB.54.4038, PhysRevLett.87.047203, TODO201984} For a recent historical review, see [\onlinecite{LajkoNuclPhys2017}].

Shortly after the formulation of this conjecture, research began on extending Haldane's work to SU($n$) generalizations of spin chains.\cite{AffleckSUn1988, Affleck1986,Affleck_1989} At this time, these were hypothetical models with no experimental realization, and their study was in part motivated by a proposed relation between nonlinear sigma models and the quantum Hall effect. \cite{PhysRevLett.51.1915, Affleck1986, RevModPhys.80.1355} While this is still a reason to study such models, recent proposals from the cold atom community suggest that SU($n$) chains may be experimentally realizable in the near future, offering a much more physical motivation.\cite{PhysRevLett.91.186402, PhysRevLett.92.170403, Cazalilla_2009, 2010NatPh...6..289G, PhysRevB.86.224409, 2014NatPh..10..779S, 2012NatPh...8..825T, 2014NatPh..10..198P, Zhang1467, Cazalilla_2014, 2016AnPhy.367...50C} These proposals have led to a renewed theoretical interest in the field of SU($n$) spin chains.\cite{GreiterRachel2007,GreiterDMRG2008, Katsura_2008, PhysRevB.80.180420, 2012PhRvB..86w5142D, Nonne_2013}

In 2017, a generalization of Haldane's conjecture to SU(3) chains was formulated.\cite{LajkoNuclPhys2017} It was shown that for chains with a rank-$p$ symmetric representation at each site of the chain (see Figure \ref{fig:young}, left), a Haldane gap above the ground state is present only when $p$ is a multiple of 3; otherwise, the chain exhibits gapless excitations. In [\onlinecite{Wamer2019}], the conjecture was further extended to self-conjugate SU(3) chains, with a $(p,p)$ representation on each site (Figure \ref{fig:young}, right). In this case, a gapped phase is always found, with spontaneously broken parity symmetry occurring only for $p$ odd. 

In this article, we generalize Haldane's conjecture to SU($n$) chains in the rank-$p$ symmetric representations (Figure \ref{fig:young}, left), following the methodology presented in [\onlinecite{LajkoNuclPhys2017}]. Our main result is the prediction of gapless excitations above the ground state when $p$ and $n$ have no common divisor greater than 1. In Section \ref{section:ham}, we introduce the SU($n$) Hamiltonian, which involves local interactions up to $(n-1)$-nearest neighbours. As we will show, these longer range interactions are necessary to stabilize the classical ground state of the chain. In Section \ref{section:exact}, we review exact results pertaining to these SU($n$) chains that support our conjecture, namely the Lieb-Shultz-Mattis-Affleck theorem\cite{LSM1961,AffleckLieb1986} and explicit AKLT-type constructions\cite{AKLT1988, GreiterRachel2007} In Section \ref{section:fw}, we carry out a flavour wave analysis, which amounts to introducing Holstein-Primakoff bosons, and performing a large-$p$ expansion. In Section \ref{section:qft}, we derive a low energy quantum field theory description of the chain, and obtain the same flavour wave velocities in a perturbative expansion. In Section \ref{section:vrg}, we use the renormalization group to argue that at low enough energies, these (distinct) flavour wave velocities may flow to a common value, so that Lorentz invariance emerges, and the field theory becomes a Lorentz-invariant flag manifold sigma model (FMSM). This FMSM description of SU($n$) chains was first derived by Bykov\cite{BYKOV2012100, bykov2013geometry}, who then fine-tuned the interactions of the SU($n$) chain to achieve a unique flavour wave velocity at the bare level. These FMSMs were also studied systematically in [\onlinecite{Tanizaki:2018xto}] and [\onlinecite{ohmori2019sigma}]. In Section \ref{section:thooft}, we relate the 't Hooft anomaly matching arguments of [\onlinecite{Tanizaki:2018xto},\onlinecite{ohmori2019sigma}] to our SU($n$) spin chain, and formulate our conjecture. In Section \ref{section:strong}, we present a strong coupling analysis of the FMSM, which further supports our claims. Section \ref{section:conclusions} contains our conclusions.

\section{Hamiltonian} \label{section:ham}

\begin{figure}[h]
\includegraphics[width = .8\textwidth]{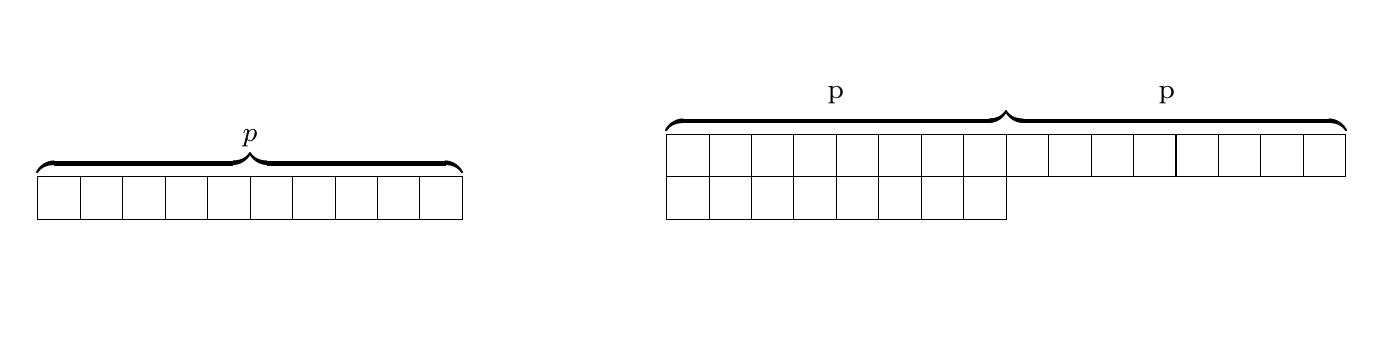}
\caption{Young tableau for irreps of SU($n$). Left: the rank-$p$ symmetric irrep considered in this paper, and also in [\onlinecite{LajkoNuclPhys2017}] for SU(3). Right: the $(p,p)$ self-conjugate irrep, considered in [\onlinecite{Wamer2019}] for SU(3).}
\label{fig:young}
\end{figure}

\begin{figure}[h]
\includegraphics[width = .7\textwidth]{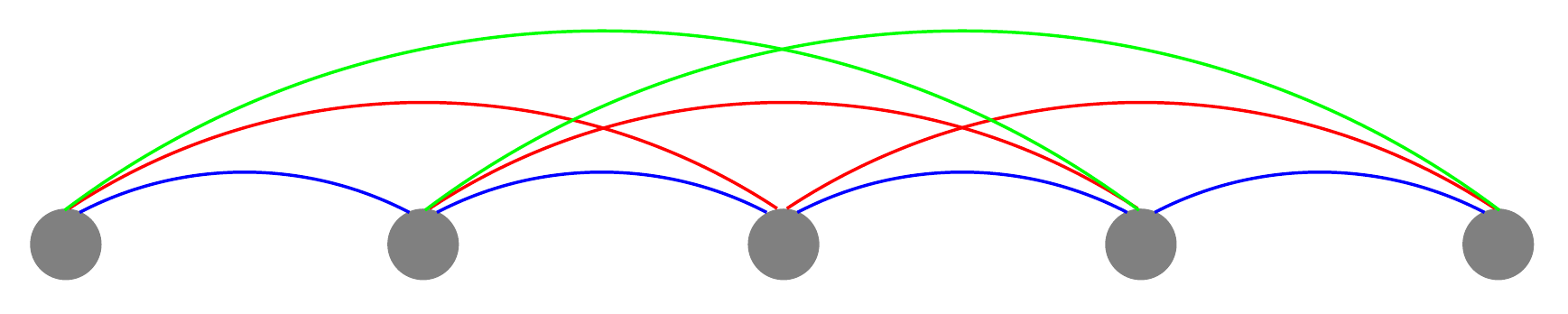}
\caption{Pictorial representation of the nearest (blue), next-nearest (red), and next-next-nearest (green) neighbour interactions occurring in (\ref{1:2}), for the case $n=4$.}
\label{cell}
\end{figure}

The familiar antiferromagnetic spin chain is characterized by a single integer, $2s$, which specifies the irreducible representation (irrep) of SU(2) that appears on each site. In SU($n$), the most generic irrep is defined by $n-1$ integers, which give the number of columns of a given length in their Young tableaux. In this paper, we focus on the rank-$p$ symmetric irreps, which have Young tableaux shown in Figure \ref{fig:young} (left). 
The simplest Hamiltonian one is tempted to write down is
\be \label{1:1}
	H = J \sum_j \tr [S(j) S(j+1)]
\ee
where $S(j)$ is an $n\times n$ Hermitian matrix with $\tr[S] =p$,\begin{footnote}{S(j) should be traceless; we have shifted it by a constant to simplify our calculations.}\end{footnote} whose entries correspond to the $n^2-1$ generators of SU($n$) and satisfy\begin{footnote}{Throughout, we use upper indices for the rows of a matrix, and lower indices for the columns of a matrix. This ordering is switched for complex conjugated entries.}\end{footnote}
\be \label{1:3}
	[S^\alpha_\beta, S^\mu_\nu] = \delta^\alpha_\nu S^\mu_\beta - \delta^\mu_\beta S^\alpha_\nu.
\ee
Indeed, in SU(2), $S^\alpha_{\beta} = \vec{S} \cdot \vec{\sigma}^\alpha_\beta + \frac{p}{2}\mathbb{I}$, and the Hamiltonian appearing in (\ref{1:1}) equals the Heisenberg model with spin $s = \frac{p}{2}$ (up to a constant). However, for $n>2$, this Hamiltonian possesses local zero mode excitations that destabilize the classical ground state and inhibit a low energy field theory description. To remedy this, we introduce an additional $n-2$ interaction terms, arriving at
\begin{equation} \label{1:2}
H = \sum_j \sum_{r=1}^{n-1} J_r \tr[ S(j)S(j+r)]
\end{equation}
where $J_1$ couples nearest-neighbours, $J_2$ couples next-nearest neighbours, and so on. See Figure~\ref{cell} for a pictorial representation of these interactions.

\subsection*{Classical Ground State} \label{sub:class}

In the large-$p$ limit, the commutator (\ref{1:3}) is subleading in $p$, allowing us to replace $S$ by a matrix of classical numbers. To this order in $p$, the Casimir constraints of SU($n$) completely determine the eigenvalues of $S$. We have 
\be \label{1:1:2}
	S^\alpha_\beta = p\phi^{*,\alpha} \phi_\beta
\ee
for $\vphi \in \mathbb{C}^n$ with $|\vphi|=1$. The interaction terms appearing in (\ref{1:1}) reduce to 
\be \label{eq:orthog}
	\tr[S(j)S(j+r)] = p^2 |\vphi(j)^* \cdot \vphi(j+r)|^2.
\ee
Since $\vphi$ lives in $\mathbb{C}^n$, a classical ground state will posses local zero modes unless the Hamiltonian gives rise to $n-1$ constraints. This is the justification for our study of the modified Hamiltonian (\ref{1:2}), above, which removes any local zero modes by including longer range interactions. These interactions result in an $n$-site ordered classical ground state, which gives rise to a $\mathbb{Z}_n$ symmetry in their low energy field theory description. This $\mathbb{Z}_n$ symmetry is also present in the $p=1$ Bethe ansatz-solvable models.\cite{PhysRevB.12.3795, doi:10.1080/00018738300101581, RevModPhys.55.331} In fact, it is expected that quantum fluctuations may produce an $n$-site unit cell through an ``order-by-disorder'' mechanism that generates effective additional couplings of order $p^{-1}$ that lift the local zero modes.\cite{LajkoNuclPhys2017, PhysRevX.2.041013}

Since the classical ground state minimizing (\ref{1:2}) has an $n$-site order, it is characterized by $n$ normalized vectors that mutually minimize (\ref{eq:orthog}). That is, the classical ground state gives rise to an orthonormal basis of $\mathbb{C}^n$. Due to this $n$-fold structure, we rewrite the Hamiltonian (\ref{1:2}) as a sum over unit cells (indexed by $j$):
\begin{equation} \label{1:1:1}
H = \sum_j \sum_{\alpha=1}^n \sum_{r=1}^{n-1} J_r \tr[ S(j_\alpha)S(j_\alpha+r)] \hspace{10mm} j_\alpha := nj + (\alpha-1)
\end{equation}

In the following sections we will expand about this classical ground state to characterize the low energy physics of (\ref{1:2}). But before this, we review some exact results that apply to SU($n$) Hamiltonians.

\section{Exact Results} \label{section:exact}

\subsection{LSMA Theorem}

\begin{figure}[h]
\includegraphics{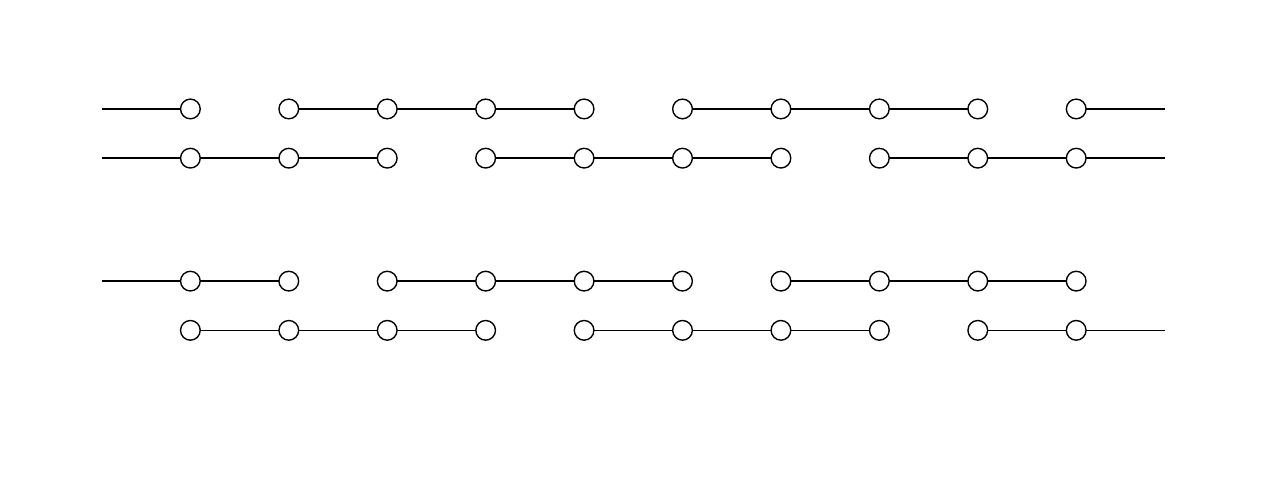}
\caption{A valence bond construction for the predicted two-fold degenerate ground state of SU(4) with $p=2$. Each node represents a fundamental $p=1$ irrep of SU(4). Each link represents an antisymmetrization between two nodes, and the antisymmetrization of four neighbouring nodes results in a singlet.}
\label{fig:su4aklt}
\end{figure}

\begin{figure}[h]
\includegraphics[width=.5\textwidth]{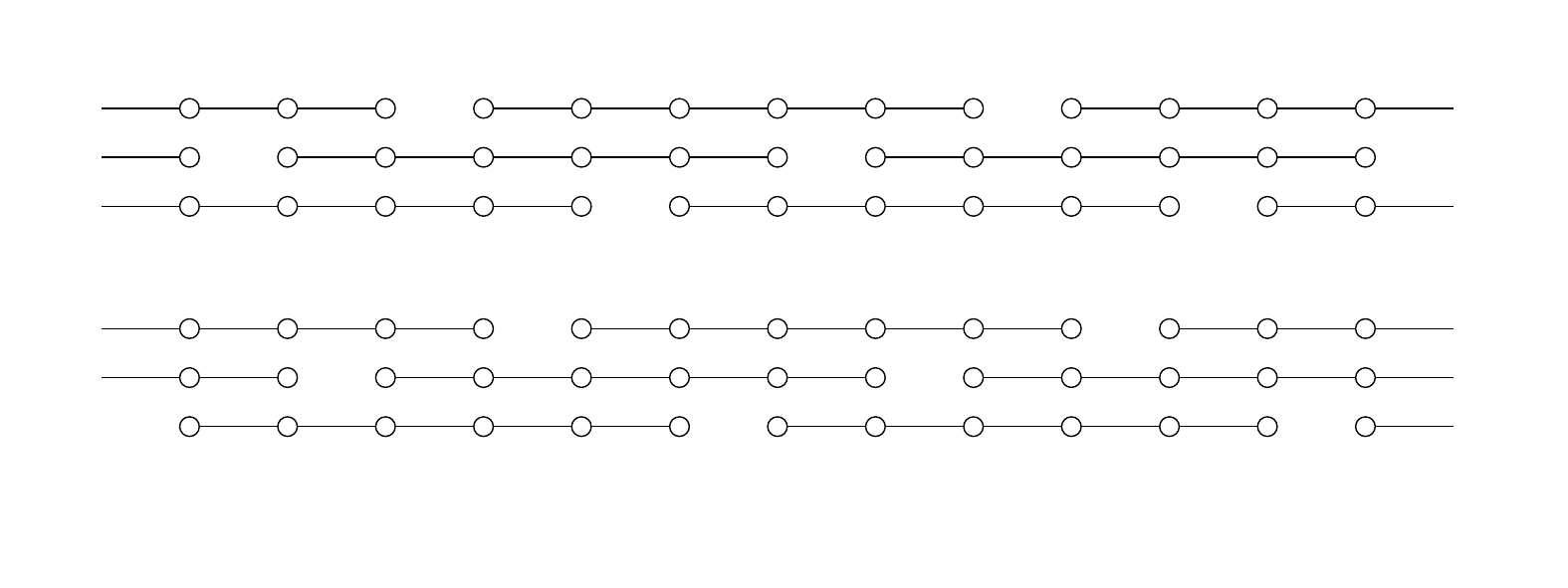}
\includegraphics[width=.49\textwidth]{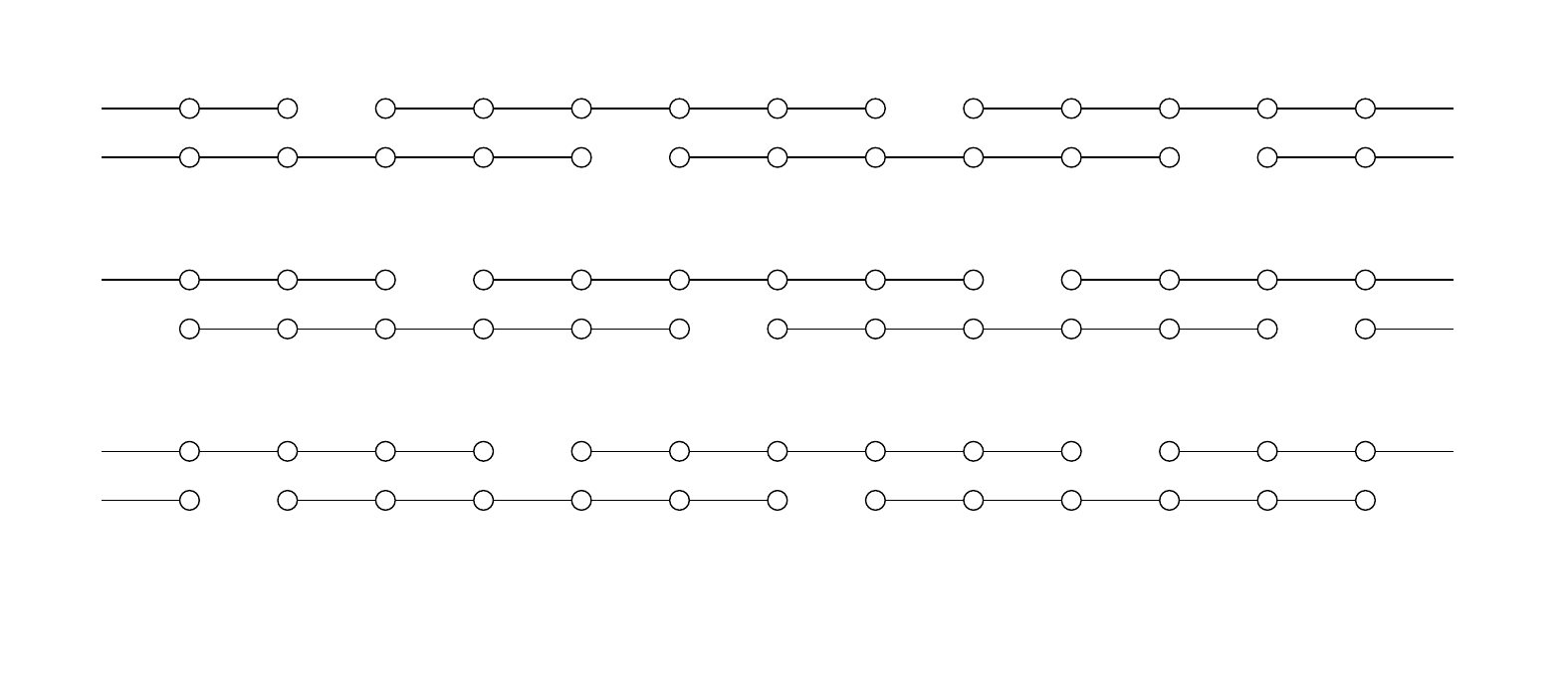}
\caption{Valence bond constructions for SU(6). The left subfigure corresponds to $p=3$, and has a 2-fold degenerate ground state. The right subfigure corresponds to $p=2$, and has a 3-fold degenerate ground state. Singlets are constructed out of 6 nodes, each of which represents a fundamental irrep in SU(6).} 
\label{fig:su6aklt}
\end{figure}

The Lieb-Schultz-Mattis-Affleck Theorem (LSMA) is a rigorous statement about ground states in translationally invariant SU($n$) Hamiltonians.\cite{LSM1961,AffleckLieb1986} Applied to the symmetric irreps considered here, this theorem proves that if $p$ is not a multiple of $n$, then either the ground state is unique with gapless excitations, or there is a ground state degeneracy. Recently, it was claimed in [\onlinecite{GreiterRachel2007}] that the LSMA theorem is not applicable to models with longer range interactions than nearest-neighbour. Here, we dispute this claim by extending the original proof in [\onlinecite{AffleckLieb1986}] to models with further range interactions. Explicitly, we consider the following Hamiltonian on a ring of $L$ sites:
\be
	H = \sum_{r=1}^R H_r \hspace{10mm} H_r :=\sum_{j=1}^L J_r \tr[S(j)S(j+r)]
\ee
where $S$ is defined as above. 
 We assume that $|\psi\rangle$ is the unique ground state of $H$, and is translationally invariant: $T|\psi\rangle = |\psi\rangle$. We then define a twist operator
\be
	U = e^A \hspace{10mm} A:= \frac{2\pi i}{n L}\sum_{j=1}^L jQ_j
\ee
with
\be \label{la:2}
	Q = \sum_{\alpha=1}^{n-1} S_\alpha^\alpha - (n-1)S^n_n
	= \tr S - nS_n^n = p - nS_n^n.
\ee
Using the commutation relations (\ref{1:3}), it is easy to verify that 
\be
	\Big[\tr[S(j)S(j+r)],Q_j+Q_{j+r}\Big] = 0
\ee
which then implies 
\be
	U^\dag \tr[S(j)S(j+r)] U = e^{-\frac{r\pi i}{n L}(Q_{j+r}-Q_j)} \tr [S(j)S(j+r)]e^{\frac{r \pi i}{nL}(Q_{j+r}-Q_j)}.
\ee
Using this, one can show that
\be
	U^\dag H U -U =[ H,A] - H + \fO(L^{-1})
\ee
so that $U|\psi\rangle$ has energy $\fO(L^{-1})$. Now, using the translational invariance of $|\psi\rangle$, we find 
\be
	\langle \psi | U | \psi\rangle
	= \langle \psi | T^{-t} UT|\psi\rangle 
	= \langle \psi| U e^{\frac{2\pi i}{n}Q_1} e^{-\frac{2\pi i}{nL}\sum_{j=1}^L Q_j} |\psi\rangle.
\ee
Since $|\psi\rangle$ is a ground state of $H$, it is a SU($n$) singlet, and so must be left unchanged by the global SU($n$) transformation $e^{-\frac{2\pi i}{nL}\sum_{j=1}^L Q_j} $. Moreover, using (\ref{la:2}), we have 
\be \label{la:4}
	\langle \psi |U|\psi\rangle = e^{\frac{2\pi i p}{n} }\langle \psi | U e^{2\pi i S_n^n} |\psi\rangle.
\ee
As we will show below, the matrices $S$ can be represented in terms of Schwinger bosons; the diagonal elements are then number operators for these bosons. Thus, $S_n^n$ acting on $|\psi\rangle$ will always return an integer, and $e^{2\pi i S_n^n}$ can be dropped. Thus, we find that so long as $p$ is not a multiple of $n$, 
\be
	\langle \psi |U|\psi\rangle = 0
\ee
implying that $U|\psi\rangle$ is a distinct, low-lying state above $|\psi\rangle$. This completes the proof. Finally, we may also comment on the ground state degeneracy in the event that a gap exists above the ground state. Through the repeated application of (\ref{la:4}), we have 
\be
	\langle \psi | U^{k} |\psi\rangle = e^{\frac{2\pi i pk}{n}} \langle \psi | U |\psi\rangle.
\ee
So long as $k < r:= n/\gcd(n,p)$, the family $\{ U^k |\psi\rangle\}$ is an orthogonal set of low lying states. If an energy gap is present, this suggests that the ground state is at least $r$-fold degenerate. See Figures \ref{fig:su4aklt} and \ref{fig:su6aklt} for a valence bond solid picture of these degeneracies in SU(4) and SU(6), respectively.

\subsection{AKLT Constructions}

\begin{figure}[h]
\includegraphics[width = .7\textwidth]{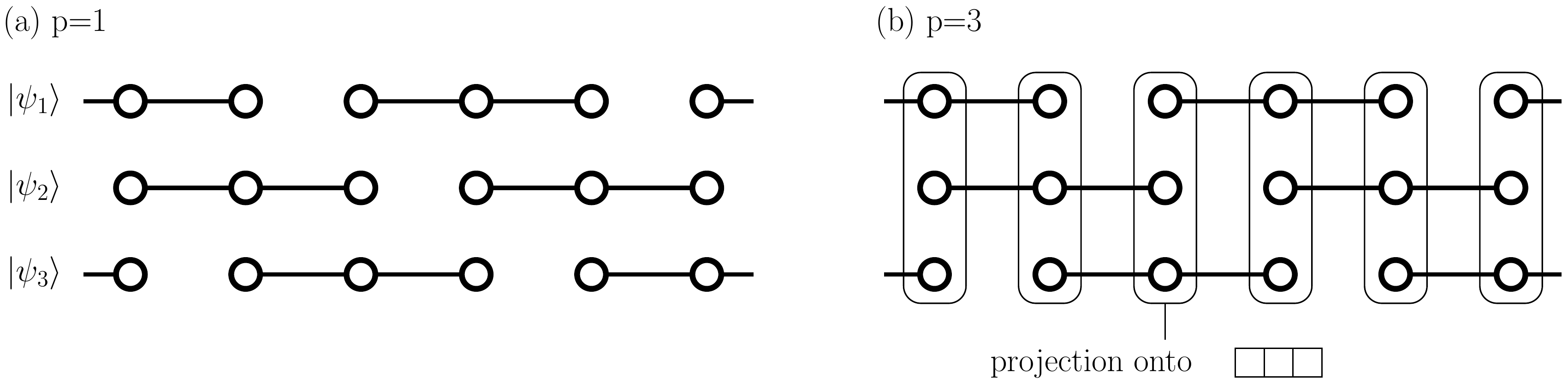}
\caption{AKLT constructions in SU(3). Left: When $p\not=n$, multiple valence bond solids can be formed. The ground state is not translationally invariant and degenerate. Right: When $p=n$, a unique, translationally invariant ground state can be constructed, by projecting on to the symmetric-$p$ representation at each site.}
\label{aklt}
\end{figure}

One of the first results that bolstered Haldane's conjecture was the discovery of the so-called AKLT model of a spin-1 chain, which exhibits a unique, translationally invariant ground state with a finite excitation gap.\cite{LSM1961,AffleckLieb1986} In this case, the number of boxes in the Young tableau is 2, and so the SU(2) version of the LSMA theorem does not apply. Recently, the AKLT construction has been generalized by various groups to SU($n$) chains.\cite{GreiterRachel2007, Katsura_2008, Nonne_2013, PhysRevB.90.235111, roy2018chiral, gozel2019novel} Relevant to us are the symmetric representation AKLT Hamiltonians introduced in [\onlinecite{GreiterRachel2007}]. In particular, for $p$ a multiple of $n$, Hamiltonians are constructed that exhibit a unique, translationally invariant ground state. See Figure \ref{aklt} for the case $n=p=3$. Additionally, for $p$ not a multiple of $n$, with $ r:= n/ \gcd(n,p)$, Hamiltonians are constructed with $r$-fold degenerate ground states that are invariant under translations by $r$ sites (see Figures \ref{fig:su4aklt}, \ref{fig:su6aklt}). All of these models have short range correlations, and are expected to have gapped ground states, based on arguments of spinon confinement. The fact that the construction of a gapped, nondegenerate ground state is only possible when $p$ is a multiple of $n$ is consistent with the LSMA theorem presented above. 

\section{Flavour Wave Theory} \label{section:fw}

According to the Mermin-Wagner-Coleman theorem\cite{MerminWagner1966, Coleman1973}, we do not expect spontaneous symmetry breaking of the SU($n$) symmetry in the exact ground state of our Hamiltonian. Nonetheless, we may still expand about the classical (symmetry broken) ground state to predict the Goldstone mode velocities. If the theory is asymptotically free, then at sufficiently high energies the excitations may propagate with these velocities. In the familiar antiferromagnet, this procedure is known as spin wave theory; in SU($n$), it is called flavour wave theory.

To begin, we introduce $n^2$ bosons in each unit cell to reproduce the commutation relations of the $S$ matrices: 
\be
	S^\alpha_\beta(j_\gamma) = b^{\dag,\alpha}(j_\gamma) b_\beta(j_\gamma).
\ee
The counting is $n$ flavours of bosons for each of the $n$ sites of a unit cell. The condition $\tr[S] = p$ implies there are $p$ bosons at each site. The classical ground state involves only `diagonal' bosons of the type $b_\gamma(j_\gamma)$ and $b^{\dag,\gamma}(j_\gamma)$.  The `off-diagonal' bosons are Holstein-Primakoff bosons. Flavour wave theory allows for a small number of Holstein-Primakoff bosons at each site, captured by
\[
	\nu(j_\gamma) = \sum_{\alpha\not=\gamma} b^{\dag,\alpha}(j_\gamma)
	 b_\alpha(j_\gamma),
\]
 and writes the Hamiltonian (\ref{1:2}) in terms of these $n(n-1)$ bosons. In the large $p \gg \nu (j_\gamma)$ limit, we expand 
\begin{align*}
S^\gamma_{\gamma} = p - \nu(j_\gamma), \\
S^\alpha_{\gamma}(j_\gamma) \approx \sqrt{p}b^{\dag,\alpha}(j_\gamma), \\
S^\gamma_{\gamma}(j_\gamma) \approx \sqrt{p}b_\alpha(j_\gamma), \\
\end{align*}
to find
\be
	\tr[S(j_\gamma) S(j_{\eta})] =
	p \left[ b^{\dag,\gamma}(j_{\eta}) b_\gamma(j_{\eta})
	+ b^{\dag,\gamma}(j_{\eta})b_{\eta}(j_\gamma)
	+ b^{\dag,\eta}(j_\gamma)b^{\dag, \gamma}(j_{\eta})
	+ b_{\eta}(j_\gamma)b_\gamma(j_{\eta})\right]
	+\fO(p^{0}).
\ee
 In terms of these degrees of freedom, the Hamiltonian (\ref{1:2}) decomposes into a sum
\be
	H = \sum_{\gamma <\eta} H_{\gamma\eta},
\ee
where $H_{\gamma\eta}$ is a Hamiltonian involving only the two boson flavours $b_\gamma(j_{\eta})$ and $b_{\eta}(j_\gamma)$. In momentum space, this gives $\frac{n(n-1)}{2}$ different $2\times 2$ matrices, each of which can be diagonalized by a Bogoliubov transformation:
\be
	H_{\gamma, \gamma+t} = \text{const.} + \sum_k \omega_t(k) \sum_{i=1}^2 \left( d^{\dag,i,t}(k)d_{i,t}(k) + \frac{1}{2}\right)
\ee
where
\be
	\omega_t(k) = 2p\sqrt{J_t J_{n-t}}\left| \sin \frac{ nka}{2}\right|.
\ee

Therefore, the corresponding flavour wave velocities are
\begin{equation}
v_t = np \sqrt{J_t J_{n-t}} \hspace{10mm} t=1,2,\cdots, n-1
\end{equation}
When $n$ is odd, there are $n$ modes with each flavour wave velocity. When $n$ is even, this is true except for the velocity $v_{\frac{n}{2}}$, which has only $\frac{n}{2}$ modes. In each case, the number of modes adds up to $n(n-1)$. We note that for $n>3$, there is no longer a unique velocity, and the emergence of Lorentz invariance is absent. Only for a specific fine tuning of the couplings can Lorentz invariance be restored. These tuned models were the ones considered by Bykov in [\onlinecite{BYKOV2012100}] and [\onlinecite{bykov2013geometry}].

\section{Derivation of Field Theory} \label{section:qft}

In SU(2), it is well known that spin wave theory fails to capture much of the low energy physics of the spin chain. In particular, it is oblivious to the presence of a topological theta term, with angle $\theta = 2\pi s$. Likewise, while flavour wave theory accurately predicts the absence of Lorentz invariance, it is incomplete, and must be supplemented with a field theoretic description. In the following section, we follow the procedure outlined in [\onlinecite{LajkoNuclPhys2017}] for deriving such a theory from an SU($n$) chain. Details of this derivation can be found in Appendix \ref{app:derivation}.

Since the classical ground state has $n$-site sublattice order, with unit vectors $ \vvphi \in \mathbb{C}^n$ defined on each site, we may defined a unitary matrix, $U$, by
\be
	U^\alpha_\beta = \varphi^\alpha_\beta.
\ee
Throughout, a superscript index labels the vector, and a subscript index labels the component of the vector (opposite for the complex conjugate vectors). To describe fluctuations about the $\vvphi^\beta$, we write a new unit vector $\vphi^\beta$ in terms of the original orthonormal basis:
\be \label{qft:l}
	\vphi^\beta =\sum_\alpha  \frac{1}{p}L^\beta_{\alpha} \vvphi^\alpha  + \sqrt{1 - \mu(\beta)} \vvphi^\beta.
\ee
Here $L^\alpha_{\alpha}=0$ (no sum) and
\be \label{qft:mu}
	p^2\mu(\beta) :=  \sum_{\alpha} |L^\alpha_{\beta}|^2.
\ee

These complex coefficients $L^\alpha_{\beta}$ describe general fluctuations about the $\vvphi^\beta$. By redefining the unitary matrix $U$, we may take $L$ to be Hermitian.\begin{footnote}{The skew components of $L$ generate unitary transformations, and can be recombined with the matrix $U$.}\end{footnote} Now, by letting $U$ and $L$ vary uniformly from site to site, and using the large $p$ expression of $S^\alpha_{\beta}(j_\gamma)$ in (\ref{1:1:2}), we show in Appendix \ref{app:derivation:1} that
\begin{equation} \label{2:1}
S(j_\gamma) 
	= p U^\dag \Lambda_\gamma U + U^\dag \{L,\Lambda_\gamma\}U +p^{-1} U^\dag\left(L\Lambda_\gamma L\Lambda_\gamma - p^2 \mu(\gamma)\Lambda_\gamma\right)U
\end{equation}

where $\Lambda_\gamma$ is zero except at entry $(\gamma,\gamma)$, where it equals 1. Using this, we now evaluate the trace terms appearing in (\ref{1:2}). Since the matrices $U$ and $L$ are evaluated at different sites, we Taylor expand which introduces spatial derivatives.  For example
\be
	U(j_\gamma) = U(nj + (\gamma-1)) = U(j_\eta)+ (\eta - \gamma)\partial_x U(j_\eta) + \frac{1}{2}(\eta-\gamma)^2\partial_{x}^2U(j_\eta) + \cdots
\ee
where we've assumed the derivative is uniform:  $\partial_x U(j_\eta) = \partial_x U(j'_\lambda)$. Expanding in powers of $L$ and $p^{-1}$, we find  
\be \label{t1t5}
	\tr[ S (j_\gamma) S (j_\eta)]
	= p^2(\eta - \gamma)^2 \tr  U\partial_x U^\dag \Lambda_\gamma \partial_x U U^\dag \Lambda_\eta
\ee
\[
	+	2(\eta-\gamma)p\left( L^\eta_{\gamma}[\partial_x U U^\dag]^\gamma_{\eta} 
	+ L^\gamma_{\eta}[U \partial_x U^\dag]^\eta_{\gamma}\right)
	+4 |L^\eta_{\gamma}|^2  + \text{const.}
\]
For the complete derivation, refer to Appendix \ref{app:derivation:1}.
\vspace{10mm}

\subsection{Coherent State Path Integral}

Having rewritten the Hamiltonian in terms of $U$ and $L$, we now derive the Lagrangian by using a coherent state path integral approach.\cite{MathurSen2001, Mathur:2002mx} As a complete set of states, we introduce

\be
	|\vphi \rangle = \phi^{\alpha_1}\phi^{\alpha_2}\cdots \phi^{\alpha_p} |\alpha_1,\alpha_2,\cdots,\alpha_p\rangle.
\ee

These states correspond to an element of the rank-$p$ symmetric irrep of SU($n$), $[{\vphi}]_p$, acting on a highest-weight state in the Hilbert space: 
\be
	|\vphi \rangle = [\vphi]_p |\text{highest weight} \rangle.
\ee

The resolution of the identity is then the integration over all of SU($n$) of the projection $|\vphi\rangle\langle \vphi|$:
\[
	1 = \int \fD \phi |\vphi \rangle \langle \vphi |.
\]

Inserting this between each time slice $\tau_i$ of the partition function, we obtain terms of the form
\[
	\langle \vphi(\tau_i) | e^{-H\delta \tau} | \vphi(\tau_{i+1}) \rangle = \langle \vphi(\tau) | \vphi_{\tau+\delta\tau}\rangle e^{-H \delta \tau }.
\]
Exponentiating these terms, we find the following contribution to the action:
 \be
	\prod_i \langle \vphi(\tau_i) | \vphi(\tau_{i+1})\rangle \propto (1 + \vphi(\tau_i)^*\cdot \partial_\tau \vphi(\tau_i) )^p	= \exp p \log \sum_i (1  + \vphi(\tau_i)^*\cdot \partial_\tau\vphi(\tau_i)) \approx \exp p\int d\tau \vphi^* \cdot \partial_\tau \vphi,
\ee
where we've used $\langle \vphi | \vphi'\rangle = (\vphi^* \cdot \vphi')^p$. Inserting (\ref{2:1}), we show in Appendix \ref{app:derivation:2} that the action receives the following `Berry' contribution:
\begin{equation} \label{topterm}
	S_{B} = -\int d\tau \left(  p\tr[\Lambda_\alpha \partial_\tau U U^\dag]
				+ \tr[ \{\Lambda_\alpha, L\} \partial_\tau U U^\dag]
			\right)
				+ O(p^{-1}).
\end{equation}

\subsection{Complete Field Theory}

Since our approximated action is only quadratic in the $L$ matrix elements, we may integrate out these modes to obtain an action in terms of the $U$ matrices only. This is done in Appendix \ref{app:derivation:3}. In the end, we obtain the following field theory describing the SU($n$) chain in the rank-$p$ symmetric irep:
\be \label{3:3}
	S =  \sum_{\alpha<\beta} \int dx d\tau \frac{1}{g_{|\alpha-\beta|}} \left( v_{
|\alpha-\beta|}\tr[\Lambda_\alpha U  \partial_xU^\dag \Lambda_\beta \partial_x U U^\dag]
	+ \frac{1}{v_{|\alpha-\beta|}} \tr[\Lambda_\alpha U \partial_\tau U^\dag \Lambda_\beta\partial_\tau U U^\dag] \right)
\ee
\[
	  -\epsilon_{\mu\nu} \sum_{\alpha < \beta} \lambda_{|\alpha-\beta|} \int dx d\tau  \tr[\partial_\mu U U^\dag \Lambda_\alpha \partial_\nu U U^\dag \Lambda_\beta]   + S_{\text{top}},
\]
where $ v_{t} = np\sqrt{J_{t}J_{n-t}}$ is the flavour wave velocity associated with the pair of couplings $J_{t}$ and $J_{n - t}$, and $S_{\text{top}}$ is a topological $\theta$-term (discussed below). The coupling constants are
\be
	g_{t} = \frac{n}{v_{t}}(J_t + J_{n-t})
\ee
and
\be \label{eq:lamform}
	\frac{ n \lambda_{t}}{p} = \frac{(n-t)J_{n-t} -tJ_{t}}{J_{t} + J_{n-t}}.
\ee
Since the coupling constants and velocities satisfy $g_t = g_{n-t}$ and $v_t = v_{n-t}$, we conclude that there are $\lfloor n \rfloor$ velocities and coupling constants, where
\be \label{eq:floor}
	\lfloor n \rfloor = \begin{cases} \frac{n}{2} & \text{$n$ even} \\
	\frac{n-1}{2} & \text{$n$ odd} \\
	\end{cases}
\ee

The topological term is 
\be 
	S_{\text{top}} := \frac{2\pi i p}{n}\sum_{\alpha=2}^{n} (\alpha-1)Q_\alpha
	\label{eq:Stop}
\ee
where
\be
Q_\alpha :=  \frac{1}{2\pi i} \epsilon_{\mu\nu} \int dx d\tau  \tr[\partial_\mu U \partial_\nu U^\dag \Lambda_{\alpha}]
\ee
is a quantized topological charge.\cite{ohmori2019sigma} Since
\be
	\sum_{\alpha=1}^n Q_\alpha  = \frac{1}{2\pi i}\epsilon_{\mu\nu} \int dx d\tau \tr[\partial_\mu U \partial_\nu U^\dag]
	= \frac{1}{2\pi i}\epsilon_{\mu\nu} \int dx d\tau \partial_\mu \tr[U \partial_\nu U^\dag] =0
\ee
we see that there are $n-1$ independent topological charges. We note that the $\lambda$-terms appearing in (\ref{3:3}) are not quantized, despite the fact that they are pure imaginary in imaginary time. We give an interpretation of these terms below. In [\onlinecite{bykov2013geometry}], these $\lambda$-terms were absent as a result of the same fine-tuning that ensured a unique velocity. Indeed, the choice $J_t = \sqrt{\frac{n-t}{t}}$ ensures that $v_t \equiv 1$ for all $t$, and moreover that $\lambda_t = 0$ for all $t$.

\subsection{Gauge Invariance}

The theory (\ref{3:3}) is invariant under the gauge transformations
\be
	U(x,\tau) \to e^{iD(x,\tau)} U(x,\tau),
\ee
where $D(x,\tau)$ is a local, diagonal matrix. Since matrices of the form $e^{iD}$ are generated by the $n-1$ diagonal SU($n$) generators, this corresponds to a $[\mbox{U}(1)]^{n-1}$ gauge symmetry. In fact, we may view $U$ as a map from (compact) spacetime $S^2$ to the flag manifold $\mbox{SU}(n)/[\mbox{U}(1)]^{n-1}$. For this reason, the above Lagrangian is known as a $\mbox{SU}(n)/[\mbox{U}(1)]^{n-1}$ flag manifold sigma model (FMSM). Since 
\be
	\pi_2(\mbox{SU}(n)/ [\mbox{U}(1)]^{n-1}) = \mathbb{Z}_{n-1},
\ee
this model is characterized by $n-1$ topological charges, which is consistent with $S_{\text{top}}$ in (\ref{3:3}). The coupling constants $\{g_t\}$ and $\{\lambda_t\}$ correspond to the metric and torsion on this manifold, respectively.\cite{ohmori2019sigma} However, a unique metric cannot be defined, since the theory (\ref{3:3}) lacks the Lorentz invariance that is often assumed for sigma models. Thus, we have a non-Lorentz invariant flag manifold sigma model, just as was the case in [\onlinecite{Wamer2019}] where self-conjugate SU(3) chains were considered. In the following section, we will use the renormalization group to show that at low enough energies, it is possible for the distinct velocities occurring in our model to flow to a single value, so that Lorentz invariance emerges.

\section{Velocity Renormalization} \label{section:vrg}

Recently, the Lorentz invariant versions of the above flag manifold sigma model were studied in great detail in [\onlinecite{ohmori2019sigma}]. In particular, the renormalization group flow of both the $\{g_{t}\}$ and $\{\lambda_{t}\}$ were determined for general $n$. Moreover, field theoretic versions of the LSMA theorem were formulated, using the methods of 't Hooft anomaly matching (which we review below, in Section \ref{section:thooft}). In this paper, we would like to apply these results to our SU($n$) chains which lack Lorentz invariance in general. To do so, we consider the differences of velocities occurring in (\ref{3:3}), namely
\be \label{delta}
	\Delta_{tt'} := v_t - v_{t'},
\ee
and ask how they behave at low energies. More precisely, we calculate the one-loop beta functions of these $\Delta_{tt'}$, to $\fO(g_{t})$ and $\fO(\lambda_{t})$. We will find that each of the $\Delta_{tt'}$ flows to zero under renormalization; moreover, we will show that this implies Lorentz invariance at our order of approximation. This is consistent with the fundamental SU($n$) models with $p=1$, where it is known by Bethe ansatz that Lorentz invariance is present.\cite{PhysRevB.12.3795, doi:10.1080/00018738300101581, RevModPhys.55.331} Our calculations were motivated by a similar phenomenon in 2+1 dimensional systems, where an interacting theory of bosons and Weyl fermions renormalizes to a Lorentz invariant model.\cite{lee2007emergence, grover2014emergent}

\subsection{Goldstone Mode Expansion}

In the following, it will be useful to introduce dimensionless velocities, $u_t$, defined according to
\be
	u_t := \frac{v_t}{\bar v} \hspace{15mm} \bar v = \frac{1}{\lfloor \frac{n}{2} \rfloor}\sum_{t=1}^{\lfloor \frac{n}{2} \rfloor} v_t,
\ee
and introduce new spacetime coordinates which both have units of $(\text{length} \cdot \text{ time})^{1/2}$:
\be
	x \to  \frac{x}{\sqrt{\bar v}} \hspace{10mm} \tau \to \sqrt{\bar v}  \tau.
\ee
 In these units, $\Delta_{tt'} = u_t-u_{t'}$. 
The coefficients $\{g_t\}$ appearing in (\ref{3:3}) are dimensionless, and are all proportional to $\frac{1}{p}$. Since we've taken a large $p$ limit, we will expand all quantities in powers of the $\{g_t\}$. As we will see below, the coefficients $\{\lambda_t\}$ in (\ref{3:3}) do not enter into our one-loop calculations, and so we will neglect them throughout.

Since we are interested in the low energy dynamics of these quantum field theories, we make the simplifying assumption that the matrices $U$ are close to the identity matrix, and expand them in terms of the SU$(n$) generators. If we use greek letters to index the diagonal generators, and lower case latin letters to index the off-diagonal generators, then it turns out that we may factorize any SU($n$) matrix $U$ according to 
\be \label{eq:factorize}
	U =DV \hspace{10mm} \begin{cases} D = e^{i\phi_\gamma T_\gamma}\\
	V =  e^{i\phi_a T_a} \\
	\end{cases}.
\ee
This is proven in Appendix \ref{app:factorize}. Since $D$ is diagonal, we see that it drops out from the traces:
\be
	\tr[U\partial_\mu U^\dag \Lambda_\alpha \partial_\mu U U^\dag \Lambda_\beta ]
	= \tr[V\partial_\mu V^\dag \Lambda_\alpha \partial_\mu V V^\dag \Lambda_\beta ]
\ee
Therefore, when deriving the Lagrangian of the $\{\theta\}$, we may write $U$ in terms of the off-diagonal generators only:
\be
	U = e^{i\theta_a T_a} = 1 + i\theta_a T_a - \frac{1}{2}\theta_a\theta_b T_a T_ b + \fO(\theta^3).
\ee
Throughout, repeated indices will be summed over. We choose a convenient normalization in which the off-diagonal generators have entries $1$ or $\pm i$, and satisfy
\be
	[T_a,T_b] = 2if_{abC}T_C.
\ee
Here and throughout, upper case latin letters are used to index the complete set of SU($n$) generators (including the diagonal ones). These generators are $n\times n$ matrices that have a very specific structure. There are $n-1$ diagonal ones, and $n(n-1)$ off-diagonal ones, that come in pairs. For each pair of integers $\{ \alpha,\beta \}$ with $\alpha,\beta =1, \cdots, n$ and $\alpha \not=\beta$, there are exactly two generators with nonzero $(\alpha,\beta)$ entries. We define $I_{\alpha\beta}$ to be the set of two indices corresponding to the SU($n$) generators with nonzero $(\alpha,\beta)$ entries. For example, in SU(3), the off-diagonal generators (in Gell-Mann's notation) are $T_1,T_2$ with nonzero entries in the $(1,2)$ positions; $T_4,T_5$ with nonzero entries in the $(1,3)$ positions; and $T_6,T_7$ with nonzero entries in the $(2,3)$ positions. Then,
\be
	I_{12} = \{1,2\} \hspace{5mm} 
	I_{13} = \{4,5\} \hspace{5mm}
	I_{23}  = \{6,7\}.
\ee
With this notation, we show in Appendix \ref{app:goldstone} that, to $\fO(\theta^4)$, 
\be \label{eq:gold1}
	-\tr[\partial_\mu U U^\dag \Lambda_\alpha \partial_\mu U U^\dag \Lambda_\beta]
	=\sum_{a\in I_{\alpha\beta} }\Big[ (\partial_\mu \theta_{a}^2)    + 2f_{bca}\partial_\mu\theta_a\partial_\mu\theta_b\theta_c
	+
	 \frac{4}{3}f_{bcE}f_{Eda}\partial_\mu\theta_{a}\partial_\mu\theta_b\theta_c\theta_d 
	+\partial_\mu\theta_e\theta_b\partial_\mu\theta_c\theta_d f_{eba}f_{cda} \Big].
\ee
To obtain the full Lagrangian, we must now sum over the possible combinations of $\alpha$ and $\beta$. Since 
\be
	\sum_{\alpha<\beta}^n\sum_{a \in I_{\alpha\beta}} := \sum_{\beta=2}^n\sum_{\alpha=1}^{\beta-1}\sum_{a \in I_{\alpha\beta}} = \sum_{a},
\ee
where $\sum_a$ again denotes a sum over all the off-diagonal generators of SU($n$),  the non-interacting Lagrangian has the form
\be
	\fL_0
	=  \frac{1}{g_a} \left[ \frac{1}{u_a} (\partial_\tau \theta_a)^2 + u_a(\partial_x\theta_a)^2\right], 
\ee
where 
\be
	g_a := g_{|\alpha-\beta|}\Big|_{I_{\alpha\beta} \ni a}
	\hspace{10mm}
	u_a := u_{|\alpha-\beta|}\Big|_{I_{\alpha\beta} \ni a}.
\ee
and again, all repeated indices are summed over. We rescale the fields according to 
\be
	\theta_a \mapsto \sqrt{\frac{g_a}{2} }\theta_{a}
\ee
to yield

\be \label{eq:2}
	\fL = \frac{1}{2}\left[ \frac{1}{u_a}(\partial_\tau \theta_a)^2 + u_a(\partial_x\theta_a)^2\right] 
	+\frac{\sqrt{g_ag_bg_c}}{\sqrt{2}}   \frac{h_a(\mu)}{g_a}  f_{bca}\partial_\mu\theta_a\partial_\mu\theta_b\theta_c
\ee
\[
\frac{\sqrt{g_bg_cg_d}}{4} \frac{  h_a(\mu)}{g_a}\Big[ \sqrt{g_e} \partial_\mu\theta_e \partial_\mu \theta_b \theta_c \theta_d f_{eca}f_{bda}
	+ \frac{4}{3} f_{bcE}f_{Eda}\sqrt{g_a}\partial_\mu\theta_{a}\partial_\mu \theta_b\theta_c\theta_d
	\Big] + \fO(\theta^5),
\]
where 
\be \label{eq:h}
	h_a(\mu) = \begin{cases} \frac{1}{u_a} & \mu = \tau \\
	u_a & \mu = x \\
	\end{cases}.
\ee

\subsection{Renormalization Group Equations}
In order to derive the renormalization group equations for the model (\ref{eq:2}), we introduce a set of renormalization coefficients, $\{Z^\mu_a\}$ and $\{Z_{\mu abcd}^e\}$, as follows. Since (\ref{eq:2}) has divergences at one-loop order, we rewrite the theory in terms of renormalized parameters, as 
\be \label{eq:2:renorm}
	\fL = \frac{1}{2}\left[ Z_a^\tau \frac{1}{u^r_a}(\partial_\tau \theta_a)^2 + u^r_aZ_a^x(\partial_x\theta_a)^2\right] +Z^{(1)}_{\mu abc}\frac{\sqrt{g_a^rg_b^rg_c^r}}{\sqrt{2}}   \frac{h^r_a(\mu)}{g^r_a}  f_{bca}\partial_\mu\theta_a\partial_\mu\theta_b\theta_c
\ee
\[
+ \frac{\sqrt{g^r_bg^r_cg^r_d}}{4} \frac{  h^r_a(\mu)}{g_a^r}\Big[Z^{(2),e}_{\mu abcd} \sqrt{g_e^r} \partial_\mu\theta_e \partial_\mu \theta_b \theta_c \theta_d f_{eca}f_{bda}
	+ Z^{(3)}_{\mu abcd}\frac{4}{3} f_{bcE}f_{Eda}\sqrt{g_a^r}\partial_\mu\theta_{a}\partial_\mu \theta_b\theta_c\theta_d
	\Big] + \fO(\theta^5).
\]
The superscripts `$r$' emphasize that the coupling constants and velocities appearing in (\ref{eq:2:renorm}) are different from those appearing in (\ref{eq:2}) (and are not indices to be summed over). Each of the renormalization coefficients has the form $Z= 1 +\delta Z$, where $\delta Z$ is a one-loop counterterm regularizing any UV divergence. In Appendix \ref{app:rgcalc}, we use dimensional regularization to calculate the $\{\delta Z\}$ at a fixed energy scale $M$. Then, by rescaling 
\be
	\theta_a \to \left( \frac{1}{Z_a^xZ_a^\tau}\right)^{1/4}
\ee
in (\ref{eq:2:renorm}), and comparing $(\partial_x\theta_a)^2$ terms in (\ref{eq:2}) and (\ref{eq:2:renorm}), we obtain the following equation for $u_a^{r}$:
\be \label{eq:vel}
	u_a = u_a^{r} \sqrt{\frac{Z_a^x }{Z^\tau_a}}.
\ee
The derivative of $u_a^r$ with respect to $\log M$, 
\be \label{eq:beta:1}
	\beta_{u_a} := \frac{d u_a^r}{d\log M},
\ee
is the so-called `beta function' of $u_a$, and describes the flow of $u_a$ as the energy scale, $M$, is changed. It is important to note that since this equation only depends on $Z^\tau_a$ and $Z_{a}^x$, we are only tasked with calculating divergences of two-point functions in our lowest order regularization scheme. In Appendix \ref{app:rgcalc}, we show that for $a \in I_{\alpha\beta}$, with $t=|\alpha-\beta|$, that
\be  \label{ztau:1}
Z_a^\tau  = 1 + \frac{M^\epsilon g_a u_a }{2\pi\epsilon} \Big(\sum_{\substack{i=1 \\ i\not=t}}^{n-1}  \frac{1}{u_i g_i}g_{|t-i|}
	- \frac{1}{3g_au_a }\big[ g_a + \frac{1}{2}\sum_c g_c\big]\Big)
\ee
and
\be  \label{zx:1}
Z_a^x = 1 + \frac{M^\epsilon g_a }{2\pi u_a \epsilon} \Big(\sum_{\substack{i=1 \\ i\not=t}}^{n-1}  \frac{u_i}{ g_i}g_{|t-i|}
	- \frac{u_a}{3g_a }\big[ g_a + \frac{1}{2}\sum_c g_c\big]\Big).
\ee
(no sum over $a$). Inserting these into (\ref{eq:beta:1}), and using (\ref{eq:vel}), we find that for $t=1,2, \cdots, q:= \lfloor \frac{n}{2}\rfloor$, 
\be \label{beta:result}
	\beta_{u_t} =\frac{u_tg_t}{4\pi} \sum_{\substack{i=1 \\ i\not=t}}^{n-1} \frac{g_{|t-i|}}{g_i}\left[ \frac{u_t}{u_i} - \frac{u_i}{u_t}\right].
\ee

\subsection{Renormalization of Velocity Differences}

We want to study the renormalization group flow of the velocity differences, $\Delta_{tt'}$ defined in (\ref{delta}). As mentioned above, the identity $u_t = u_{n-t}$ reduces the number of independent velocities to $q = \lfloor \frac{n}{2}\rfloor$, and the relation 
\be
	\Delta_{tt'} = \Delta_{t1} + \Delta_{1t'} = \Delta_{1t'} - \Delta_{1t}
\ee
shows that the number of independent velocity differences to $q-1$. To study their flow collectively, we introduce a $q-1$-component vector, $\bDelta$, with components
\be \label{eq:vec:del}
	\Delta^i := \Delta_{1,i+1} \hspace{10mm} i=1,2,\cdots, q-1.
\ee
If we assume that the velocities $\{u_t\}$ are initially close together, so that the SU($n$) chain is approximately Lorentz invariant, the vector $\bDelta$ will obey an equation of the form
\be \label{beta:matrix}
	\frac{d}{d\log M} \bDelta = R\bDelta
\ee
for a $(q-1)\times (q-1)$ matrix $R$. The spectrum of $R$ will reveal the low energy behaviour of the $\Delta_{tt'}$: if the spectrum is strictly positive, we may conclude that all velocity differences flow to zero in the IR. In Appendix \ref{app:beta:diff}, we provide the formulae for $R$ up to $\fO(\Delta)$. These equations are quite formidable, and we cannot treat them analytically in general. We consider some special cases, including the highly symmetric point when all of the coupling constants $\{g_t\}$ are equal. In this case, we find
\be
		R = \frac{g\Delta^t}{2\pi}(n-1)\mathbb{I}_{q-1}
\ee
showing that the spectrum of $R$ is strictly positive. Moreover, for $n=4,5,6$, we verify explicitly that the spectrum of $R$ is strictly positive. Also in Appendix \ref{app:beta:diff}, we discuss our (simplistic) numerical checks that suggest the spectrum of $R$ is strictly positive for $n\leq 50$. Based on these results,  \textbf{we conjecture that the spectrum of $R$ is strictly positive for all $\boldsymbol{n}$}, so that at low enough energies, all of the flavour wave velocities flow to a common value if they are initially close to the same value.

So far, we have verified that the velocity differences $\Delta_{tt'}$ in our FMSMs flow to zero at low energies. However, we now claim that this is sufficient to conclude that the entire theory (\ref{3:3}) is Lorentz invariant at low energies. We note that we are not required to restore the pure-imaginary $\lambda$-terms occurring in (\ref{3:3}), since they are proportional to $\epsilon_{\mu\nu}\partial_\mu\phi \partial_\nu\phi$, a Lorentz scalar. Indeed, since the interaction vertex receives no $\fO(g)$ correction, the only spacetime-dependence enters through the renormalization of $\{h_{t}(\mu)\}$ and through the renormalization of the fields $\{\theta_a\}$ themselves. Since the latter are independent of $\mu$ (see (\ref{ztau:1}) and (\ref{zx:1})), the Lorentz non-invariance of the interactions is entirely captured by the $\{h_{t}(\mu)\}$. Since $u_t -u_t' \to 0$ implies $u_t^{-1} - u_{t'}^{-1} \to 0$ at $\fO(g)$, we may use the results of the previous subsection to conclude that the $\{h_{t}(\mu)\}$ all flow to a common value $\{h(\mu)\}$, and thus Lorentz invariance of the entire model (\ref{3:3}) is possible if the velocities are initially close to each other.

\section{Flag Manifold Sigma Models and $\mbox{'t}$ Hooft Anomaly Matching} \label{section:thooft}

Based on the renormalization group analysis in the previous section, we now argue that at low enough energies, the SU($n$) chains in the symmetric-$p$ irreps (without fine-tuning), may be described by a Lorentz invariant flag manifold sigma model
\be \label{eq:rel1}
	\fL = \sum_{\alpha<\beta} \frac{1}{g_{|\alpha-\beta|}} \tr[\Lambda_\alpha U\partial_\mu U^\dag\Lambda_\beta\partial_\mu U U^\dag]
	- \epsilon_{\mu\nu}\sum_{\alpha<\beta} \lambda_{|\alpha-\beta|}\tr[\partial_\mu U U^\dag \Lambda_\alpha \partial_\nu U U^\dag \Lambda_\beta]  
\ee
with topological theta-term
\be \label{eq:fm1}
	S_{\text{top}} = i\theta \sum_{\alpha=1}^{n-1}\alpha Q_\alpha \hspace{10mm} \theta := \frac{2\pi p}{n}.
\ee
These sigma models have been studied recently.\cite{BYKOV2012100, bykov2013geometry,ohmori2019sigma} In [\onlinecite{ohmori2019sigma}], the renormalization group flow of the $\{\lambda\}$ and the $\{g\}$ were determined, and given a geometric interpretation. It was found that for $n>4$, the $\{g\}$ flow to a common value in the IR, and that for $n>6$, the $\{\lambda\}$ flow to zero in the IR. Thus we may expect an $S_n$ (permutation group) symmetry to emerge at low enough energies, and for $n>6$. It is known that in these $S_n$-symmetric models, the unique coupling constant $g$ obeys\cite{ohmori2019sigma}
\be
	\frac{dg}{d\log M} = \frac{n +2}{4\pi}g.
\ee
and the theory is asymptotically free. 

\subsection*{'t Hooft Anomaly Matching}

Using the notion of 't Hooft anomaly matching, both [\onlinecite{ohmori2019sigma}] and [\onlinecite{Tanizaki:2018xto}] were able to formulate a field-theoretic version of the LSMA theorem for SU($n$) chains. In short, the presence of an 't Hooft anomaly signifies nontrivial low energy physics; in one-dimension, this necessitates a gapless phase so long as the symmetries of the SU($n$) chain are not spontaneously broken. It was shown that in these models, an 't Hooft anomaly is present so long as $p$ is not a multiple of $n$. Explicitly, it is a mixed anomaly between the PSU($n) := \mbox{SU}(n)/\mathbb{Z}_n$ spin symmetry and the $\mathbb{Z}_n$ translation symmetry of the $n$-site-ordered classical ground state. It is a PSU($n$) symmetry, and not a SU($n$) symmetry, because of a $\mathbb{Z}_n$ subgroup of SU($n$) that acts trivially on each term in the field theory. When this anomaly is present, the gapped phase must have spontaneously broken translation or PSU($n$) symmetry; the latter is ruled out by the Mermin-Wagner-Coleman theorem at any finite temperature. In the gapped phase, the ground state degeneracy is predicted to be $\frac{n}{\gcd(n,p)}$, which is consistent with the LSMA theorem presented above.\cite{Yao:2018kel} It is interesting to note that when the classical ground state has a different structure, as in the ground state of the two-site-ordered self-conjugate SU(3) chains,\cite{Wamer2019} no anomaly occurs. This is consistent with the fact that the proof of the LSMA theorem also fails for such representations.

The authors of [\onlinecite{ohmori2019sigma}] then argued that while an anomaly is present whenever $p \mod n \not=0$, an RG flow to an IR stable WZW fixed point is possible only when $p$ and $n$ have no nontrivial common divisor. In this case, the flow is to SU($n)_1$. Otherwise, the candidate IR fixed point is SU($n)_q$, where $q=\gcd(n,p)$, which is unstable and requires fine-tuning in order for the flag manifold sigma model to flow there. Also, there is no possible flow from this unstable theory to SU($n)_1$, since this would violate the anomaly matching conditions derived in [\onlinecite{Yao:2018kel}] for generic SU($n$) WZW models.

\begin{figure}[h]
\includegraphics{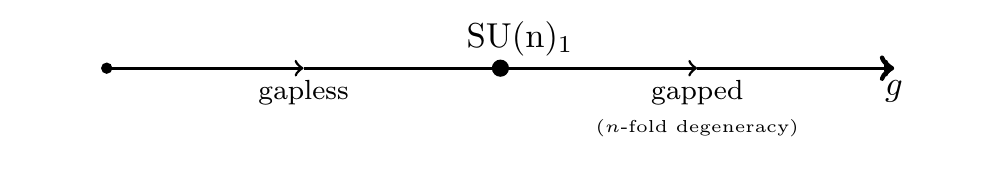}
\caption{A simplified phase diagram of the SU($n$) chains we consider, as a function of coupling constant $g$ when $p$ and $n$ are coprime. A more accurate diagram would include $\lfloor \frac{n}{2}\rfloor$ different coupling constants -- in this case, we predict a critical point described by SU($n)_1$ occurring somewhere in this multidimensional space.}
\label{fig:phasediag}
\end{figure}

Based on these anomaly arguments, we conclude that the rank-$p$ symmetric SU($n$) chains may flow to a SU($n)_1$ WZW model if $p$ and $n$ do not have a common divisor. In this case, we expect gapless excitations to appear in the excitation spectrum. This prediction is a natural extension of the phase diagrams occurring in [\onlinecite{AffleckHaldane1987}] and [\onlinecite{LajkoNuclPhys2017}]. 
We note that when $p$ and $n$ have a common divisor, at least one of the topological angles occurring in (\ref{eq:fm1}) is necessarily trivial. In the instanton gas picture of Haldane's conjecture, each type of topological excitation must have a nontrivial topological angle in order to ensure total destructive interference in half odd integer spin chains.\cite{Affleck:1985jy} This might lead one to speculate that a similar mechanism is at play here in SU($n$) chains. See Figure \ref{fig:phasediag} for a simplified phase diagram of the SU($n$) chain in the case when $p$ and $n$ are coprime. 

\section{Strong Coupling Analysis} \label{section:strong}

As it was discussed in [\onlinecite{LajkoNuclPhys2017}], and also in [\onlinecite{1997PhRvD..55.3966P}], in the strong coupling limit of (\ref{eq:rel1}), only the $\lambda$-terms and topological terms survive. We will neglect the $\lambda$-terms -- for $n>6$, this is justified by the fact that the $\lambda$ parameters flows to zero under renormalization; for $n\leq 6$, this is an added assumption that is made to simplify our analysis. In this case, with only the topological terms remaining, the path integral over the fields $U$ can be rewritten as an integral over the topological charge densities, with the extra constraints that the total topological charges take integer values, and that the topological charge densities on each plaquette of the lattice sum to zero. 
\begin{equation}
\begin{split}
Z(\theta_1, &\theta_2, \dots_, \theta_n)\\
= \prod\limits_{\vec r}& \left[ \int \limits_{-1/2}^{1/2}\int \limits_{-1/2}^{1/2} \dots \int \limits_{-1/2}^{1/2} dq_1(\vec r) dq_2(\vec r) \dots dq_n(\vec r)  \delta(q_1(\vec r)+q_2(\vec r) + \dots + q_n(\vec r))\right]  \exp \Big[ i \sum_{\vec r} \big(\theta_1 q_1(\vec r) +\theta_2 q_2(\vec r) + \dots + \theta_n q_n (\vec r)\big) \Big]\\  
&\sum_{Q_2}\delta \Bigg( \sum_{\vec r} q_2(\vec r) -Q_2\Bigg) \sum_{Q_3}\delta\Bigg( \sum_{\vec r} q_3(\vec r) -Q_3\Bigg) \dots \sum_{Q_n}\delta \Bigg( \sum_{\vec r} q_n(\vec r) -Q_n\Bigg).
\end{split}
\end{equation}
Note that we don't have a Dirac term for $\sum _{\vec r } q_1(\vec r)$, because by fixing all the other topological charges to $ Q_2, Q_3 \dots, Q_n$, the $Q_1$ is uniquely determined due to the fact the topological charges densities sum up to 0.  Using the Fourier transform of the Dirac comb
\begin{equation}
\begin{split}
\sum_{Q \in \mathbb{Z}} \delta(x-Q) =\sum_{m \in \mathbb{Z}} \exp( i2\pi m x),
\end{split}
\end{equation}
we get
\begin{equation}
\begin{split}
Z(\theta_1, \theta_2, \dots , \theta_n) =\sum_{\{m_2, \dots m_n\}} z( \theta_1 , \theta_2+2\pi m_2, \dots, \theta_n+2\pi m_n)^V,
\end{split}
\label{eq:Zsum}
\end{equation}
where $V$ is the spacetime volume and 
\begin{equation}
\begin{split}
z(\theta_1, \theta_2,\dots, \theta_n) = \int \limits_{-1/2}^{1/2}\int \limits_{-1/2}^{1/2} \dots \int \limits_{-1/2}^{1/2} dq_1 dq_2 \dots dq_n \delta(q_1+q_2 + \dots + q_n)  \exp \Big[ i \big(\theta_1 q_1 +\theta_2 q_2 + \dots + \theta_n q_n\big) \Big].
\end{split}
\label{eq:smallz}
\end{equation}


Using the Fourier transform of the Dirac-$\delta$ function, and switching the order of integration, we have
\be \label{st:3}
	z(\theta_1,\theta_2,\cdots,\theta_n) = \int_{-\infty}^{\infty} \frac{dk}{2\pi}\prod_{\alpha=1}^n \frac{2\sin \left(\frac{1}{2}(k-\theta_\alpha)\right)}{(k-\theta_\alpha)}.
\ee
We assume that the topological angles are ordered and all different: $\theta_1<\theta_2<\cdots <\theta_n$. In Appendix \ref{app:strong}, we use the method of contour integration to evaluate (\ref{st:3}), which depends on the parity of $n$. 
\subsection*{Case 1: $n$ odd}
\be \label{st:result1}
	z(\theta_1,\theta_2,\cdots,\theta_n) =
	 \sum_{\beta=1}^n \sum_{\substack{ \{s_1,\cdots, s_{\beta-1}, s_{\beta+1},\cdots, s_n\} \\
	 \sum_{\alpha\not=\beta} s_\alpha =0}} \frac{ \cos\left( \frac{1}{2} \sum_{\alpha\not=\beta} s_\alpha\theta_\alpha\right)}{\prod_{\alpha\not=\beta}^n (\theta_\beta-\theta_\alpha)},
\ee
where the second sum is over all sets $\{s_1, \cdots, s_{\beta-1}, s_{\beta+1},\cdots, s_n \}$ with $s_i \in \{-1, +1\}$, that satisfy $
	\sum_{\alpha\not=\beta}^n s_\alpha =0.$ 
\subsection*{Case 2: $n$ even}
\be \label{st:result2}
	z(\theta_1,\theta_2,\cdots,\theta_n) =
	 \sum_{\substack{ \{s_1,s_2,\cdots, s_n\} \\
	 \sum_{\alpha} s_\alpha =0}} \sin\left( \frac{1}{2}\sum_\alpha s_\alpha\theta_\alpha\right) \left[ \sum_{\substack{\beta: \\
	s_\beta = 1 }} \frac{1}{\prod_{\alpha\not=\beta}^n(\theta_\beta-\theta_\alpha)}\right].
\ee
where the first sum is over all sets $\{s_1,s_2,\cdots ,s_n\}$ with $s_i\in \{-1, +1\}$ that satisfy $\sum_\alpha s_\alpha =0$. 

\subsection{Phase transitions and degeneracies}

In the following we will use both symmetry arguments and numerical calculations based on the actual form of the partition function to explore the phase diagram in the strong coupling limit. According to Eq.~\eqref{eq:Zsum} the partition function in the strong coupling limit reads as
\begin{equation*}
\begin{split}
Z(\theta_1, \theta_2, \dots , \theta_n) =\sum_{\{m_2, m_3,\dots, m_n\}} z( \theta_1 , \theta_2+2\pi m_2, \dots, \theta_n+2\pi m_n)^V,
\end{split}
\end{equation*}
where the term(s) with the largest value dominate the sum, and therefore the free energy density reads as
\begin{equation}
\begin{split}
f(\theta_1,\theta_2, \dots, \theta_n) = -\log \Big(\max\limits_{\{m_2,m_3,\dots, m_n\}} z(\theta_1, \theta_2 +2 \pi m_2, \dots, \theta_n +2\pi m_n ) \Big).
\end{split}
\end{equation}
Using the notation 
\begin{equation}
\begin{split}
z_{\{m_2,m_3,\dots, m_n\} } (\theta_1, \theta_2, \dots, \theta_n) \equiv  z(\theta_1, \theta_2 +2 \pi m_2, \dots, \theta_n +2\pi m_n ),
\end{split}
\end{equation}
for the versions of $z(\theta_1,\theta_2,\dots, \theta_n)$, the  partition function reads as 
\begin{equation}
\begin{split}
Z(\theta_1, \theta_2, \dots , \theta_n) =\sum_{\{m_2, m_3,\dots, m_n\}} z_{\{m_2, m_3,\dots, m_n\}}( \theta_1 , \theta_2, \dots, \theta_n)^V.
\end{split}
\end{equation}
The region where the $z_{\{m_2,\dots m_n\}}$ term dominates the sum will be called the $\mathcal{R}_{\{m_2,m_3,\dots,m_n\}}$ sector.

It is easy to see that the global maximum of $z(\theta_1,\dots,\theta_n)$ is at $\theta_1=\theta_2=\dots \theta_n=0$ since at this point the integrand in Eq.~\eqref{eq:smallz} is identically 1. So at  $(\theta_1, \theta_2,\dots \theta_n)=(0,0,\dots,0)$   $z_{\{0,0,\dots 0\}}$ dominates  partition function. We will call this the {\it center} of the $\mathcal{R}(0,0,\dots,0)$ sector. Similarly, the center of $\mathcal{R}_{\{m_2,m_3,\dots,m_n\}}$ will be at $\theta_1=0, \theta_2=-2\pi m_2, \dots, \theta_n =-2\pi m_n$. At the  boundary between different sectors there are multiple equally large terms in the sum in the partition function. When we cross the boundary, the dominant term will change and therefore the derivative of the free energy will have a cusp, indicating  a first order phase transition.  The number of sectors meeting at a given point will give the degeneracy at the transition. In the following we will use symmetry arguments to locate possible phase transitions, and then we will compare our findings to actual numerical results. 

Shifting all angles by the same value $\Delta$ will leave $z$ unchanged, as will any permutation of the topological angles. Therefore we have
\begin{equation}
\begin{split}
z( \theta_1,\theta_2,\dots, \theta_n) = z( \theta_{\perm{1}}+\Delta,\theta_{\perm{2}}+\Delta,\dots, \theta_{\perm{n}}+\Delta)
\end{split}
\end{equation}
for any permutation $\perm{} \in S_n$. We further note that changing the sign of all angles simultaneously is also a symmetry, so 
\begin{equation}
\begin{split}
z( \theta_1,\theta_2,\dots, \theta_n) = z( -\theta_{\perm{1}}+\Delta,-\theta_{\perm{2}}+\Delta,\dots,- \theta_{\perm{n}}+\Delta).
\end{split}
\end{equation}
is also true for any permutation $\perm{}$.  Based on these  we can formulate a condition for two terms $z_{\{m_2,m_3,\dots, m_n\}}$ and  $z_{\{n_2,n_3,\dots, n_n\}}$ to be degenerate at a given $(\theta_1, \theta_2, \dots,\theta_n)$ point. By definition they are equal if 
\begin{equation}
\begin{split}
  z(\theta_1,\theta_2+2\pi\, m_2,\dots,\theta_n+2\pi\, m_n)= z(\theta_1,\theta_2+2\pi\, n_2,\dots,\theta_n+2\pi\, n_n),
\end{split}
\end{equation}
which can happen if the angles on the right hand side are a permutation of the ones on the left  (up to a constant shift), i.e.\
\begin{equation}
\begin{split}
(\theta_{\perm{1}} + 2 \pi\,m_{\perm{1}}, \theta_{\perm{2}}+ 2\pi\, m_{\perm{2}},\dots, \theta_{\perm{n}}+2\pi\, m_{\perm{n}} ) =\mathbin{\phantom{-}}   (\theta_{1} +2\pi\, l_1, \theta_{2}+ 2\pi \,l_2 \dots, \theta_{n}+2\pi \, l_n )  +\Delta,
\end{split}
\label{eq:degcond}
\end{equation}
or if
\begin{equation}
\begin{split}
(\theta_{\perm{1}} + 2 \pi\,m_{\perm{1}}, \theta_{\perm{2}}+ 2\pi\, m_{\perm{2}},\dots, \theta_{\perm{n}},+2\pi\, m_{\perm{n}} ) =   -(\theta_{1} +2\pi\, l_1, \theta_{2}+ 2\pi \,l_2 \dots, \theta_{n}+2\pi \, l_n )  +\Delta,
\end{split}
\label{eq:degcondv2}
\end{equation}
where on the left hand side  $m_1= 0$. 

In the following we will focus on the  case when  $\theta_\alpha = \theta (\alpha-1)$, and in particular on the specific points with  $\theta =2\pi p/n$ that correspond to translational invariant chains of spins in the fully symmetric p-box irrep. For these points the cyclic permutations $\perm{\alpha}^{(k)}= \alpha + k \,(\text{mod }n)$, with $k=1,2,\dots,  n-1$ always give a solution of Eq.~\eqref{eq:degcond}, for any given $\{m_2,m_3,\dots,m_n\}$,

\begin{equation}
\begin{split}
\def\arraystretch{1.5}
\left(\begin{array}{c}
\frac{2k\pi p}{n} + 2\pi m_{k+1}\\
\frac{2(k+1)\pi p}{n} +2\pi m_{k+2}\\
\frac{2(k+2)\pi p}{n} +2\pi m_{k+3}\\
\vdots\\
\frac{2(n-1)\pi p}{n} +2 \pi m_{n}\\
0 \\ 
\frac{2 \pi p}{n} + 2\pi m_2\\
\vdots\\
\frac{2(k-1)\pi p}{n} + 2\pi m_{k}
 \end{array} \right)  =
 \left(\begin{array}{c}
0\\
\frac{2\pi p}{n} +2\pi l_2\\
\frac{4\pi p}{n} +2\pi l_3\\
\vdots\\
\frac{2(n-k-1)\pi p}{n}+ 2\pi l_{n-k}\\
\frac{2(n-k)\pi p}{n}+ 2\pi l_{n-k+1}\\
\frac{2(n-k+1)\pi p}{n}+ 2\pi l_{n-k+2}\\
\vdots\\
\frac{2(n-1)\pi p}{n}+2\pi l_n
\end{array} \right)+\Delta^{(k)},
\end{split}
\end{equation}
which gives
\begin{equation}
\begin{split}
\Delta^{(k)}&= \frac{2 \pi p k}{n}+ 2\pi m_{k+1},\\
l_2&= m_{k+2} - m_{k+1},\\
l_3&= m_{k+3} - m_{k+1},\\
&\vdots\\
l_{n-k} &= m_n -m_{k+1},\\
l_{n-k+1} &= m_1 - m_{k+1} - p,\\
l_{n-k+2} &= m_2- m_{k+2} - p,\\
&\vdots\\
l_n & = m_{k}-m_{k+1}-p.
\end{split}
\label{eq:condsolution}
\end{equation}
where we included $m_1=0$ to better show the pattern.
If $n$ and $p$ are coprime, each $\perm{}^{(k)}$ will give a different $\{l_2, l_3,\dots,l_n\}$ solution. By summing the $n$ equations,  we can verify that two different cyclic permutations $\perm{}^{(k)}$ and $\perm{}^{(k')}$ can only give the same solution for the $l_\alpha$s if $\Delta^{(k')} = \Delta^{(k)}$. This cannot happen if $n$ and $p$ are coprime because $\Delta^{(k')}- \Delta^{(k)} \neq 0 (\text{mod } 2\pi)$ no matter what are the values of $m_k, m_{k'}$. For the same reason each $\{l_2,l_3,\dots, l_n\}$ solution is different from $\{m_2,m_3,\dots, m_n\}$ as well.   So for any given $z_{\{m_2,m_3,\dots, m_n\}}$, we always find $n-1$ other degenerate terms in the partition function, resulting in a  degeneracy that has to be a multiple of $n$, as long as $\gcd(n,p)=1$. For the actual form of the partition function for small $n$ we always find a degeneracy $n$ when $\gcd(n,p)=1$.  This argument is analogous to the argument of Refs.~[\onlinecite{ohmori2019sigma}] and [\onlinecite{Tanizaki:2018xto}], using the anomaly due to the $\mathbb{Z}_n$ symmetry present at the $\theta_\alpha =2\pi p(\alpha-1)/n$ points.

If $\gcd(n,p)=q>1$ not all $\perm{}^{(k)}$ give a necessarily different solution for a given $\{m_2,m_3,\dots,m_n\}$. A  $\perm{}^{(k)}$ can give a trivial solution  of Eq.~\eqref{eq:degcond} (i.e.\ where $m_\alpha = l_\alpha$) if 
$\Delta^{(k)}=0$, which is only possible if k is a multiple of $ r= n/q$. In this case we can still use the previous arguments to show that  $\perm{}^{(1)}, \perm{}^{(2)},\dots \perm{}^{(r-1)}$ give different solutions for any $\{m_2,m_3,\dots, m_n\}$. But, for example, if a term $\{m_2,m_3,\dots, m_n\}$ satisfies 
\begin{equation}
\begin{split}
\frac{2 \alpha \pi p}{n}  +2\pi m_{\alpha+1} = \frac{2 (\alpha+r) \pi p}{n}  +2\pi m_{\alpha+r+1},
\end{split}
\label{eq:rcond}
\end{equation}
for all $\alpha$, then the $\perm{}^{(r)}$ cyclic permutation gives the trivial solution $m_\alpha=l_\alpha$, and the permutation $\perm{}^{(r+k)}$  will also give the same solutions as the $\perm{}^{(k)}$, so we will only get $r$ distinct sectors, giving  an $r$-fold degeneracy.   For the actual form of the partition function for small $n$ this is indeed the case, and  we typically find a degeneracy of $r=n/\gcd(n,p)$. 

\subsection{The case of $\theta<2\pi/n$}

As we mentioned before, for $\theta=0$ the $z_{\{0,0,\dots,0\}}$ term dominates the partition function, and here we show  that the symmetry arguments predicts  it  to remain non-degenerate for all $\theta < 2\pi/n$.  This  can be easily seen by looking at Eqs.~\eqref{eq:degcond} and \eqref{eq:degcondv2} for a general permutation $\perm{}$ and  $\{m_2, m_3,\dots, m_n\} =\{0,0,\dots,0\}$:
\begin{equation}
\begin{split}
 \big(({\perm{1}}-1)\cdot \theta, (\perm{2} -1)\cdot\theta, \dots, ({\perm{n}}-1) \cdot\theta \big)&= \big(0,\theta +2\pi l_2, 2\theta +2\pi l_3,\dots, (n-1)\theta + 2\pi l_n\big) +\Delta\\
  \big(({\perm{1}}-1)\cdot \theta, (\perm{2} -1)\cdot\theta, \dots, ({\perm{n}}-1) \cdot\theta \big)&= -\big(0,\theta +2\pi l_2, 2\theta +2\pi l_3,\dots, (n-1)\theta + 2\pi l_n\big)  +\Delta\\
\end{split}
\end{equation}
Taking the difference of the  $(\alpha+1)$st and  $\alpha$th terms on both sides we get  
\begin{equation}
\begin{split}
\pm (\perm{\alpha+1}-\perm{\alpha})\theta = \theta + 2\pi (l_{\alpha+1} -l_\alpha)
\end{split}
\label{eq:diffcond}
\end{equation}

The right hand side is between $-(n-1) \theta$ and $(n-1) \theta$, so if $\theta<2\pi/n$ this only has a trivial solution $l_\alpha=0$.  Since at  $\theta=0$ the $z_{\{0,0,\dots,0\}}$ is the unique dominant term in the partition function, there is no symmetry required degeneracy until $\theta=2\pi/n$.  Note that there could be ${\it accidental}$ degeneracies, i.e. degeneracies that are not due to any symmetry of the $z(\theta_1,\theta_2,\dots,\theta_n)$. However, considering  the actual form of $z(\theta_1,\theta_2,\dots,\theta_n)$ for finite $n$, we find that indeed $z_{\{0,0,\dots,0\}}$ is the unique dominant term until $\theta=2\pi/n$.  
At $\theta=2\pi/n$, considering $\{m_2,m_3,\dots,m_n\}=\{0,0,\dots,0\}$  Eq.~\eqref{eq:condsolution} gives the nontrivial solutions  $\{l_2,l_3,\dots, l_n\}= \{0,0,\dots, 0,-1\}$,  $\{0,0,\dots, 0,-1,-1\}$,\dots,$ \{-1,\dots, -1\}$, corresponding to an $n$-fold degeneracy. 

Similar arguments can be made for the $2(n-1)\pi/n<\theta< 2\pi$ interval, where the $z_{\{ -1,-2,\dots,-(n-1)\}}$ term dominates the partition function.  At $\theta=2(n-1)\pi/n$, this becomes degenerate with the $z_{\{-1,-2,\dots,-(n-2),-(n-2)\}}$,  $z_{\{-1,-2,\dots,-(n-3),-(n-3), -(n-2)\}}$, \dots,  $z_{\{-0,-1,\dots,-(n-2)\}}$  terms, giving the $n$-fold degeneracy.

\subsection{Examples}

Here we show results of the free energy density for SU(3)to SU(6), along the  $\theta_\alpha= \theta(\alpha-1)$ line which connects the points corresponding to various p-box fully symmetric irreps. The results fully agree with the prediction of the LSMA theorem and the symmetry arguments above.

\subsubsection{$\mbox{SU}(3)$}

\begin{figure}[htbp]
\begin{center}
\includegraphics[width=0.35\textwidth]{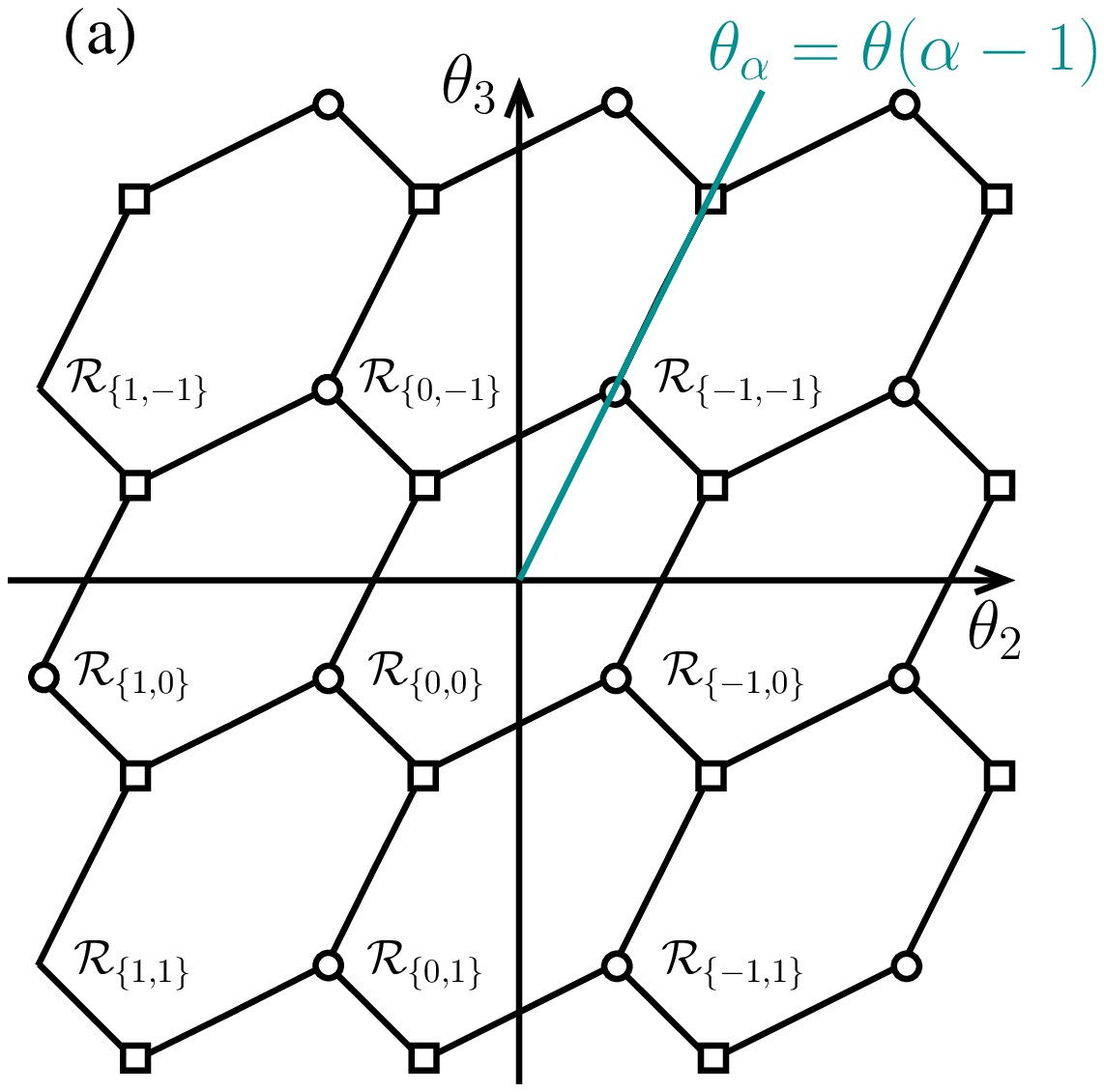}
\includegraphics[width=0.4\textwidth]{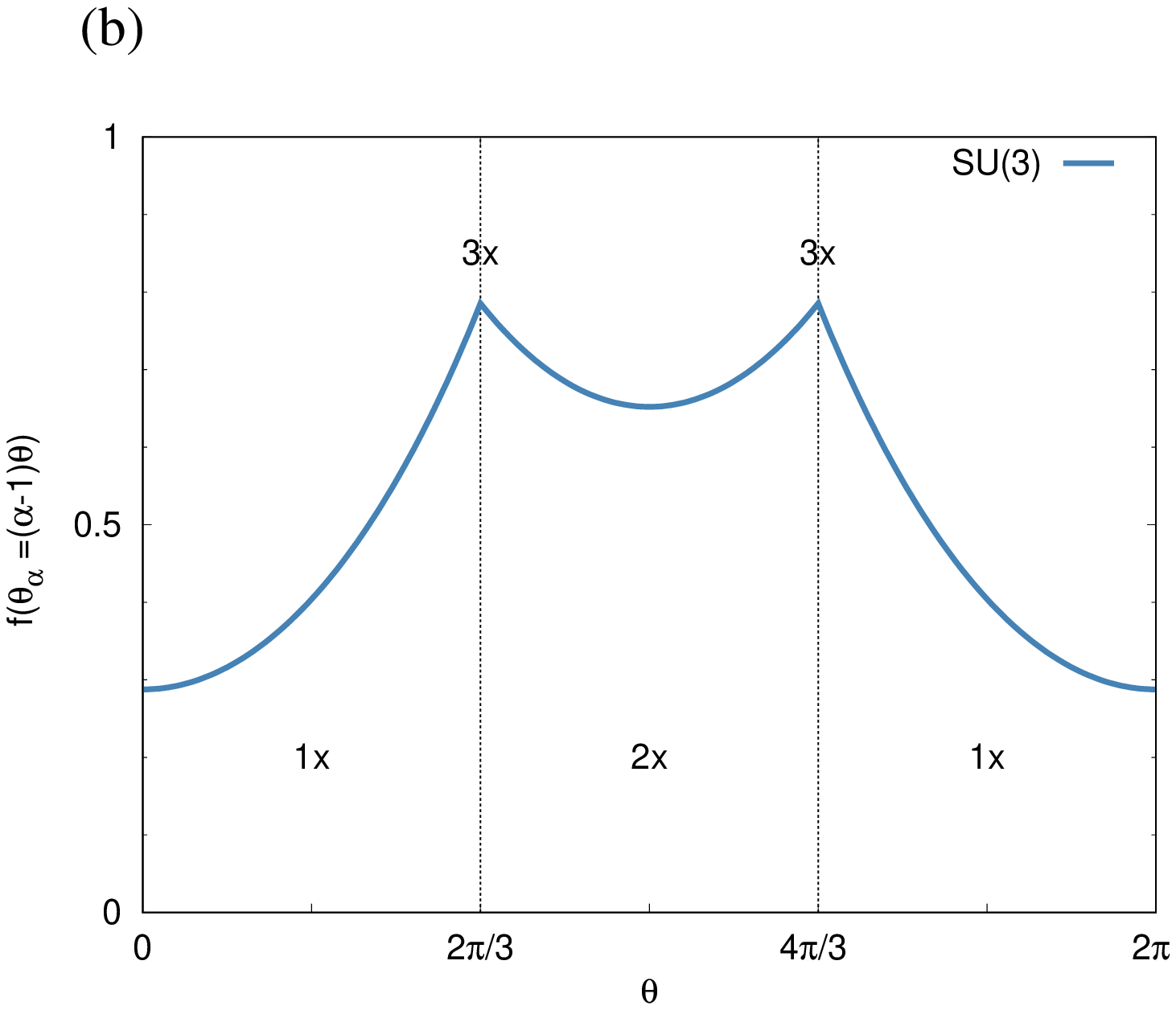}
\caption{ (a) Phase diagram of the SU(3) model, with $\theta_1=0$ fixed, highlighting the different sectors and the special $\theta_\alpha = \theta (\alpha-1)$ line. (b) The free energy density along the $\theta_\alpha = \theta (\alpha-1)$ line, showing the degeneracies in the different regions.   }
\label{fig:su3}
\end{center}
\end{figure}

 Results for $\mbox{SU}(3)$ were already presented in [\onlinecite{LajkoNuclPhys2017}], here we give a short overview, and discuss the transitions from the point of view of the symmetry arguments. Note that here we use a different convention and set $\theta_1=0$, while in [\onlinecite{LajkoNuclPhys2017}], $\theta_2$ was set to 0.  The different  $\mathcal{R}_{\{m_2, m_3\}}$ sectors are shown on Fig.~\ref{fig:su3}a, while the free energy is depicted along the $\theta_\alpha=\theta(\alpha-1)$ line on Fig.~\ref{fig:su3}b. As discussed before, starting from $\theta=0$, the first phase transition takes place at $\theta=2\pi/3$, where sectors $\mathcal{R}_{\{0,0\}}$, $\mathcal{R}_{\{0,-1\}}$, $\mathcal{R}_{\{-1,-1\}}$ meet, resulting in a threefold degeneracy. For $2\pi/3< \theta <4\pi/3$, the degenerate $z_{\{0,-1\}}$, $z_{\{-1,-1\}}$  sectors dominate the partition function. These two terms are degenerate for any $\theta$ along the $\theta_\alpha= \theta(\alpha-1)$ line, which can be seen from Eq.~\eqref{eq:degcondv2} using the permutation that exchanges $\theta_1$ and $\theta_3$, but they are  dominant  only in the $2\pi/3< \theta <4\pi/3$ interval.
Note however that not all terms in the partition function have  degenerate pairs along this line -- for example the $z_{\{0,0\}}$ is not degenerate with any other terms for general $\theta$-- so it is possible that for some exotic form of $z$, there would be no degeneracy for $2\pi/3<\theta<4\pi/3$. A similar argument was also made on the basis of anomaly and global inconsistency matching in Ref.~[\onlinecite{Tanizaki:2018xto}] (see also Ref.~[\onlinecite{Hongo2019}] for further evidence in support of the SU(3) phase diagram, which comes from considering the sigma model on $\mathbb{R}\times S^1$ with twisted boundary conditions). Note that the threefold degeneracy at $ \theta= 2\pi  /3$ and $\theta=4\pi/3$ is always present independently of the form of $z$ or of the dominant terms. 

\subsubsection{$\mbox{SU}(4)$}
 
\begin{figure}[htbp]
\begin{center}
\includegraphics[width=0.3\textwidth]{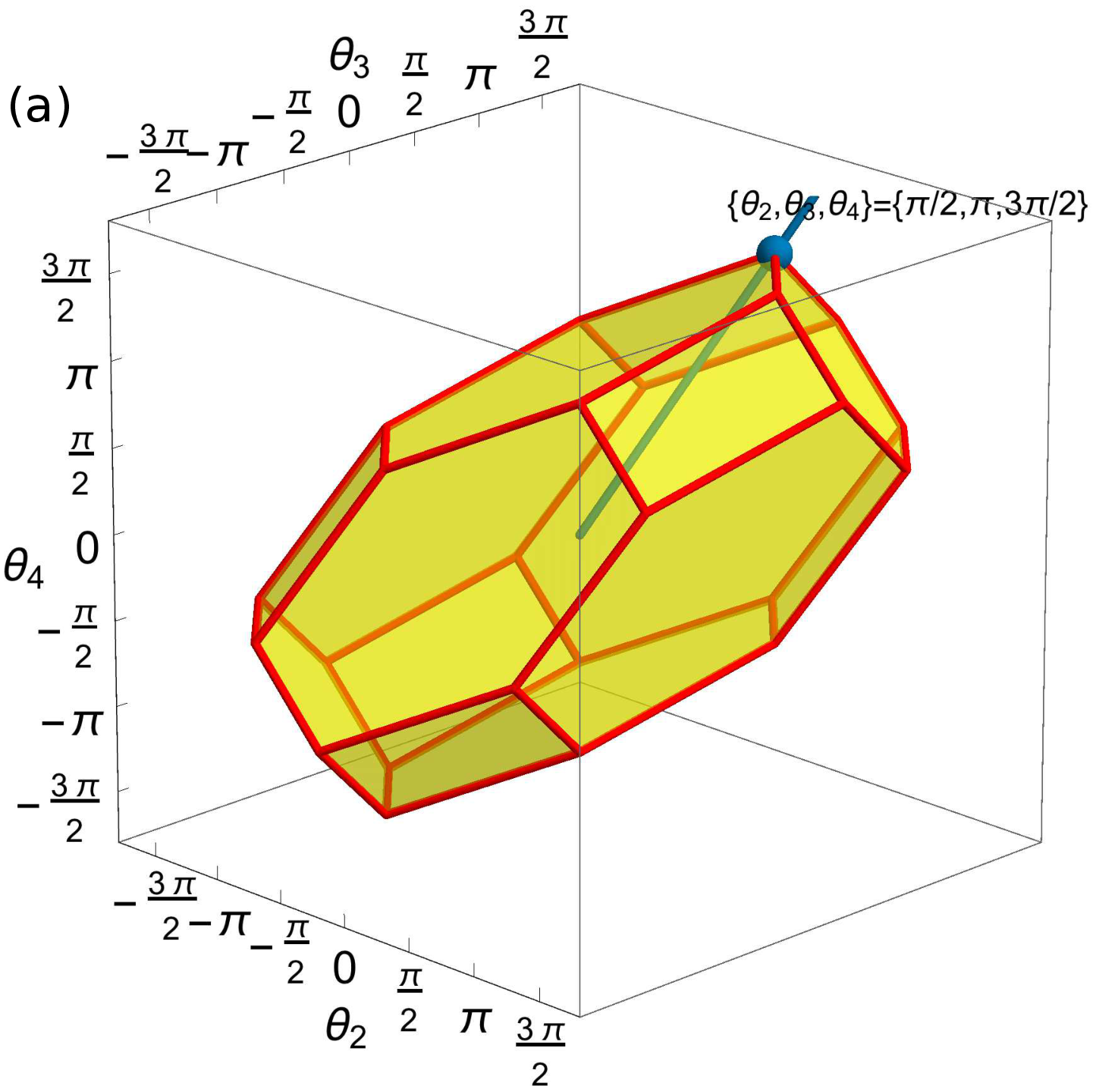}
\includegraphics[width=0.3\textwidth]{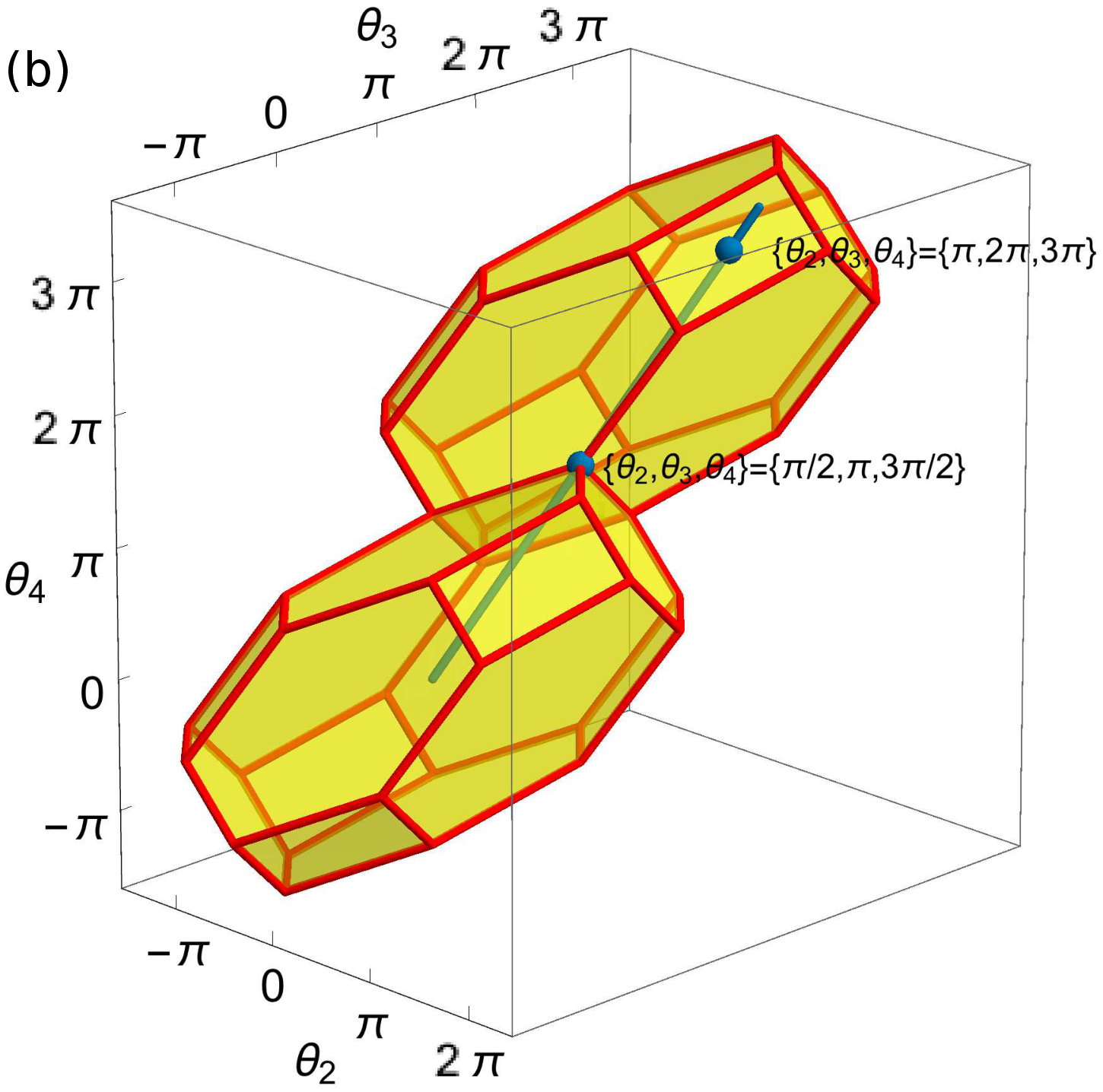}
\includegraphics[width=0.34\textwidth]{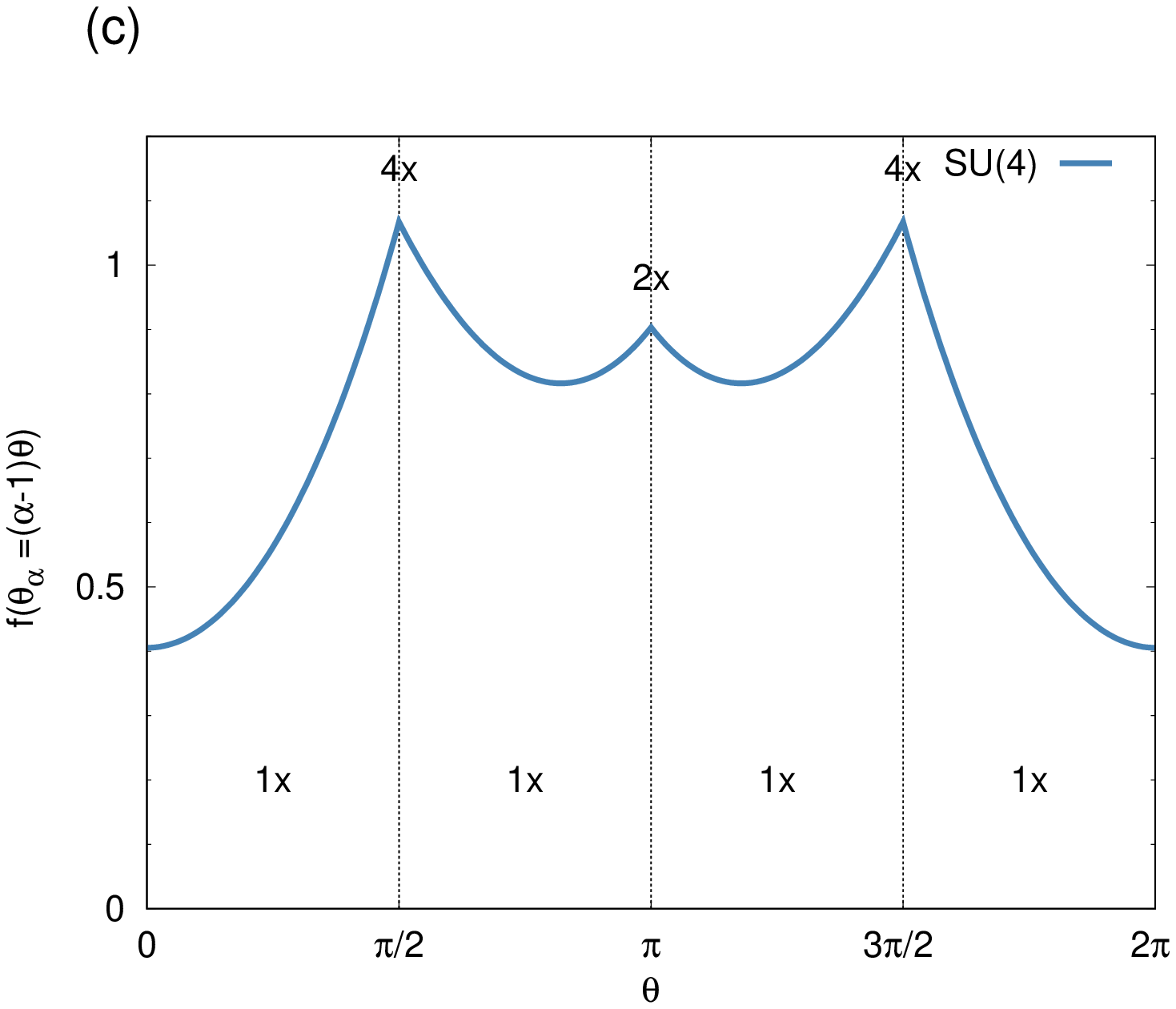}
\caption{ (a) The $\mathcal{R}_{\{0,0,0\}}$ sector for SU(4) case, highlighting  the $\theta_\alpha = \theta (\alpha-1)$ line and the $\theta=2\pi/4$ point in particular. (b) The $\mathcal{R}_{\{0,0,0\}}$ and $\mathcal{R}_{\{0,-1,-1\}}$ sectors with the $\theta_\alpha = \theta (\alpha-1)$ line and the  $\theta=\pi/2$ and  $\theta=\pi/$ points.    (c) The free energy density along the $\theta_\alpha = \theta (\alpha-1)$ line, showing the degeneracies in the different regions. }
\label{fig:su4}
\end{center}
\end{figure}

By fixing $\theta_1=0$, we can still plot the phase diagram of the $\mbox{SU}(4)$ case. In Fig.~\ref{fig:su4}a, we depict the $\mathcal{R}_{\{0,0,0\}}$ sector, which has corners that are the permutations of the  $(\theta_1,\theta_2, \theta_3,\theta_4)=(0,\pi/2,\pi,3\pi/2)$ point.  In Fig.~\ref{fig:su4}b we show a neighboring sector, which also contains the  $(\theta_1,\theta_2, \theta_3,\theta_4)=(0,\pi,2\pi,3\pi)$, clearly showing that at that point only two sectors meet.   In Fig.~\ref{fig:su4}c  we show the free energy density along the $\theta_\alpha= \theta(\alpha-1)$ line together with the degeneracies. Once again the degeneracies at the $\theta=2\pi p/4$ points are the same as predicted by  the symmetry arguments. Between $\pi/2 <\theta < \pi$ the $z_{\{0,-1,-1\}}$ term dominates the free energy, while between $\pi/2 <\theta < \pi$ it is the $z_{\{-1,-1,-2\}}$. At $\theta=\pi$ these two terms give the 2-fold degeneracy. 

\subsubsection{$\mbox{SU}(5)$}

\begin{figure}[htbp]
\begin{center}
\includegraphics[width=0.45\textwidth]{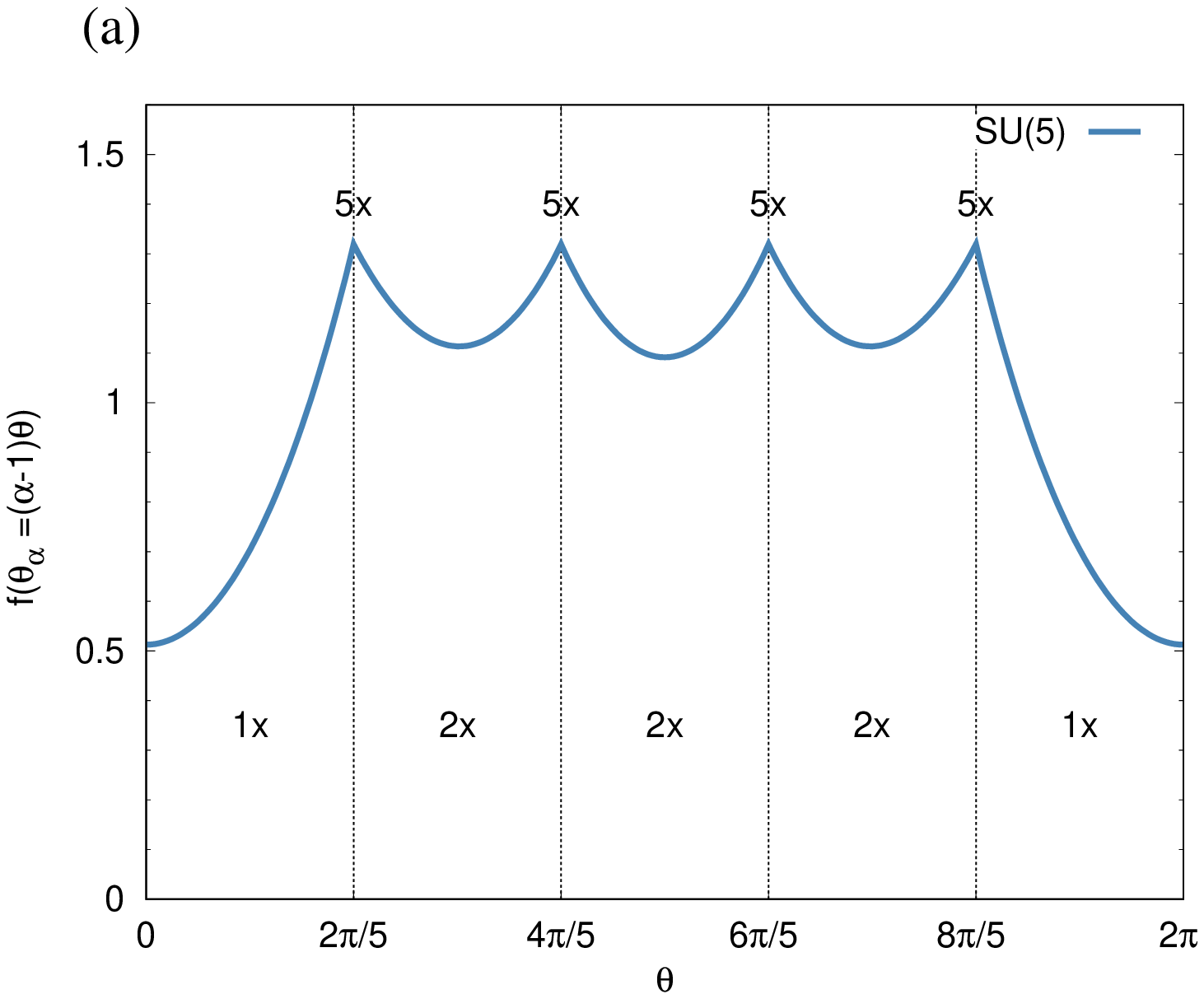}
\includegraphics[width=0.45\textwidth]{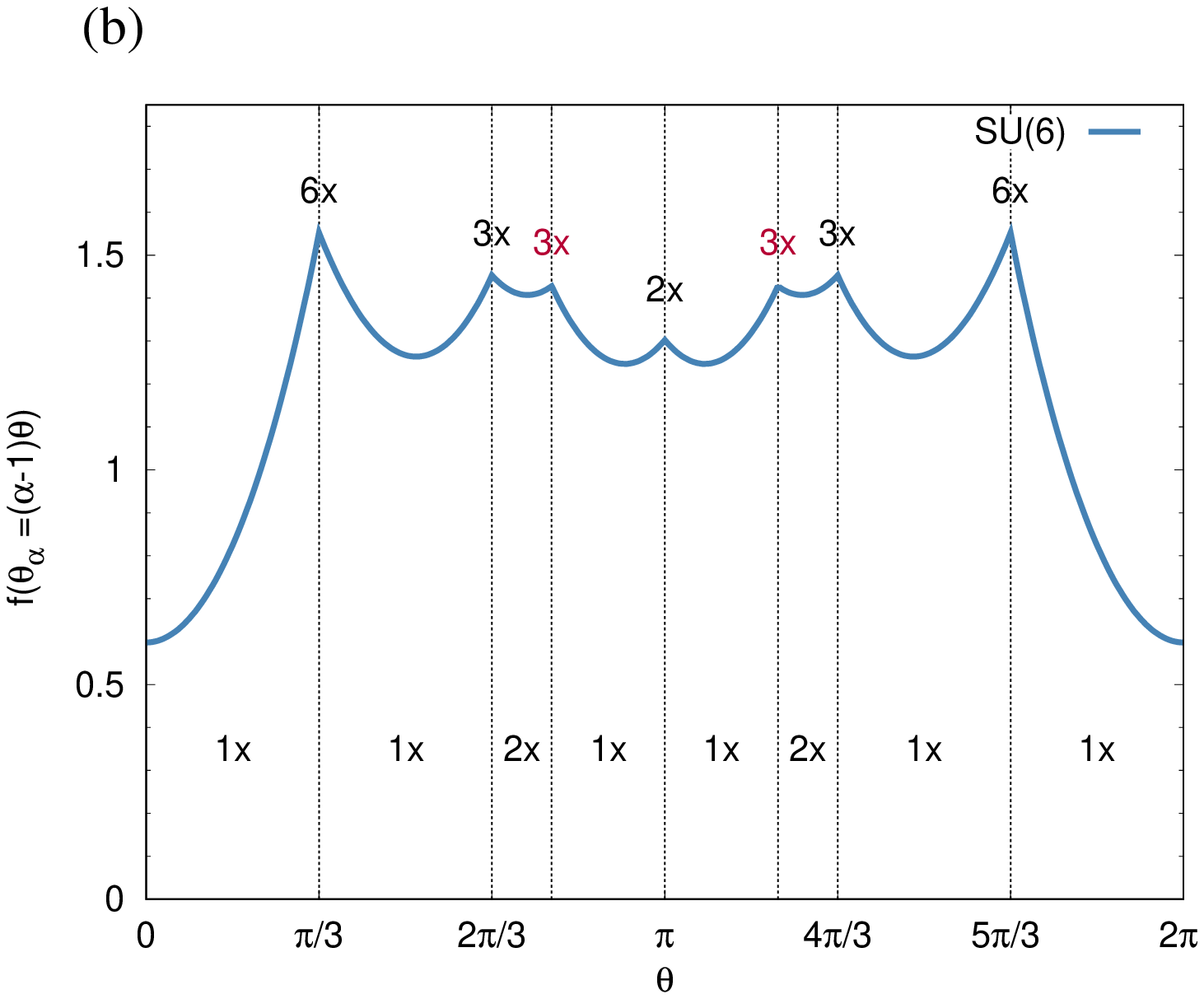}
\caption{Free energy density along the $\theta_\alpha=\theta(\alpha-1)$ line for SU(5) (a) and SU(6) case (b). We highlight the degeneracies in the different regions and at the transition points. In the case of  $\mbox{SU}(6)$,  we also highlight the unexpected transitions at the $\theta\approx 2.45629$ and at $\theta= 3.82689$.  }
\label{fig:su56}
\end{center}
\end{figure}

In the case of $\mbox{SU}(5)$ all $\theta =2\pi p/5$ points are fivefold degenerate as expected, while in the intervals in between, the system is twofold degenerate. Similarly to the SU(3) case this degeneracy is due to the actual form of  $z$, and could be removed if a different sector is dominant in this interval.  In Fig. \ref{fig:su56}a, we show the free energy density together with the degeneracies along the $\theta_\alpha =\theta(\alpha-1)$ line. At the $\theta=2\pi p/5$ points we find fivefold degenerate transition points. In the intervals $0<\theta<2\pi/5$ and $8\pi/5<\theta<2\pi$ the system is trivial, while in the other intervals $2\pi p/5<\theta<2\pi(p+1)/5$ the phases are twofold degenerate. For any prime $n$, the free energy should have a similar form, with $n$-fold degenerate transition points at $\theta = 2\pi p/n$ and $2$-fold degeneracy in between, except for $\theta<2\pi/n$ and $\theta > 2\pi(n-1)/n$, where the system is trivial. More generally, the number of $n$-fold degenerate points in the free energy is $\phi(n)-1$, where $\phi(n)$ is the Euler totient function, which counts he number of integers $1\leq k \leq n$ with $\gcd(k,n)=1$.

\subsubsection{$\mbox{SU}(6)$}

In the $\mbox{SU}(6)$ case, the free energy density presented in Fig.~\ref{fig:su56}b shows some unexpected features.  For the $\theta=2\pi p/6 $ points we find phase transitions with the expected degeneracies: for $p=1,5$ the system is sixfold degenerate, while for $p=2,4$ and $p=3$ it is  three- and  twofold degenerate, respectively. However, we find two other transition points at $\theta\approx 2.45629$ and $\theta= 3.82689$, where three sectors meet, and a transition takes place from a twofold degenerate phase to a trivial phase. Interestingly, the location of this transition point is not fixed by any symmetry.  Take for example the $\theta\approx 2.45629$ point. For $\theta< 2.45629$
the twofold degeneracy is the result of the meeting of the $\mathcal{R}_{\{-1,-1,-1,-2,-2\}}$ and the $\mathcal{R}_{\{0,-1,-1,-1,-2\}}$ sectors, their degeneracy is explained by Eq.~\eqref{eq:degcondv2}, taking the permutation $\perm{}$ that reverses the order of the topological angles,  $(\theta_1,\theta_2, \dots,\theta_6) \to (\theta_6, \theta_5, \dots,\theta_1)$. For $\theta > 2.45629$, however, it is the $z_{\{0,-1,-1,-2,-2\}}$ term that alone dominates the partition function. The location of the transition between these two phases is neither fixed by Eq.~\eqref{eq:degcond} nor by Eq.~\eqref{eq:degcondv2}; it is an accidental degeneracy. By symmetry a similar transition takes place at $\theta \approx 3.82689$. Note that while these symmetries are not predicted by the symmetry considerations, they can by expected by the degeneracies. For $\theta \gtrsim 2\pi/3$ the system is twofold degenerate, if there were no additional transition between $2\pi/3$ and $\pi$ (nor between $\pi$ and $4\pi/3$), then at $\theta=\pi$ we would have 2+2 sectors meeting resulting in a fourfold degeneracy at least. As a result, small perturbations that preserve the symmetries can tune the location of these unexpected transitions, but they cannot be removed unless they are merged either with the transition at $\theta =\pi$ or with the ones at $\theta= 2\pi/3$ and $\theta=4\pi/3$.

\section{Conclusions} \label{section:conclusions}

In this paper, a low energy field theory was derived for SU($n$) chains in the rank-$p$ symmetric irrep, in the limit of large $p$. Using the renormalization group, it was shown that this field theory may flow to a Lorentz invariant flag manifold sigma model at low energies, a model that was recently studied in great detail by Ohmori et. al. in [\onlinecite{ohmori2019sigma}]. Based on the 't Hooft anomaly matching conditions in [\onlinecite{ohmori2019sigma}] and [\onlinecite{Yao:2018kel}], as well as the LSMA theorem, generalized AKLT constructions and a strong coupling analysis, we proposed the following generalization of Haldane's conjecture to SU($n$) chains: When $p$ is an integer multiple of $n$, the corresponding chain is in a gapped phase with a unique ground state. When $p$ is not a multiple of $n$ but $\gcd(p,n)>1$, a gapped phase is also present, but the ground state is degenerate, with degeneracy $n/\gcd(p,n)$. Finally, when $p$ and $n$ have no common divisor, there are gapless excitations above the ground state, with a critical point described by an SU($n)_1$ WZW model. For interaction strengths greater than this critical coupling, spontaneously broken $\mathbb{Z}_n$ symmetry is predicted, with an $n$-fold degenerate ground state. Numerical verification of this conjecture remains a major open challenge.

\section{Acknowledgements}

KW is supported by an NSERC PGS-D Scholarship, as well as the Stewart Blusson Quantum Matter Institute's QuEST Program. IA is supported by NSERC of Canada Discovery
Grant 04033-2016 and by the Canadian Institute for Advanced Research. FM and ML are supported by the Swiss National Science Foundation. We would like to thank Sven Bachmann, Chensghu Li, Karlo Penc and Nathan Seiberg for helpful discussions.

\appendix

\section{Details of Field Theory Derivation} \label{app:derivation}

\subsection{Hamiltonian} \label{app:derivation:1}

In this appendix, we provide a detailed derivation of the field theory appearing in Section \ref{section:qft}. Our starting point is (\ref{qft:l}), which is reproduced here for convenience:
\be \label{qft:l2}
	\vphi^\alpha =\sum_\beta  \frac{1}{p}L^\alpha_{\beta} \vvphi^\beta  + \sqrt{1 - \mu(\alpha)} \vvphi^\alpha.
\ee
The matrix $L$ is off-diagonal and Hermitian, and $\mu(\alpha)$ is defined in (\ref{qft:mu}). Note that both $\vvphi$ and $L$ are allowed to vary from site to site, which is a slightly different approach than the one used in [\onlinecite{LajkoNuclPhys2017}]. Since $\vphi^*_\alpha$ is labelled by a lower index, our notation leads to $(L^\alpha_\beta)^* = L^\alpha_\beta$ and $(U^\alpha_\beta)^* = [U^\dag]^\alpha_\beta$. Using (\ref{qft:l2}), we write 
\be \label{qft:l3}
	S^\alpha_\beta(j_\gamma) = p \phi^{*,\alpha}_\gamma(j_\gamma) \phi^\gamma_\beta(j_\gamma)
	=p \sum_{\delta,\sigma} \tilde L^{\delta}_{ \gamma}(j_\gamma) U^{\dag,\alpha}_{\delta}(j_\gamma)\tilde L^\gamma_{\sigma}(j_\gamma) U^\sigma_{\beta}(j_\gamma)
\ee
where we've defined $\tilde L^\alpha_{\beta} = \frac{1}{p}L^\alpha_{\beta}$ when $\alpha\not=\beta$, and $\tilde L^\alpha_{\alpha} = \sqrt{1-\mu(\alpha)}$. This can be rewritten as
\be
	S^\alpha_\beta(j_\gamma) = p\tilde L^\gamma_{\gamma}(j_\gamma)\sum_{\delta\not=\gamma}\left( \tilde L^\delta_\gamma(j_\gamma) U^{\dag, \alpha}_\delta(j_\gamma) U^\gamma_\beta(j_\gamma)
	+U^{\dag, \alpha}_\gamma(j_\gamma)\tilde L^\gamma_\delta(j_\gamma) U^\delta_\beta(j_\gamma)\right)
\ee
\[
	+ p\left[\tilde L^\gamma_\gamma(j_\gamma)\right]^2U^{\dag,\alpha}_\gamma(j_\gamma)U^\gamma_\beta(j_\gamma)
	+ p\sum_{\delta,\sigma \not=\gamma} \tilde L^{\delta}_\gamma(j_\gamma) U^{\dag,\alpha}_{\delta}(j_\gamma) \tilde L^\gamma_{\sigma}(j_\gamma) U^{\sigma}_\beta(j_\gamma).
\]
Using
\be
	\tilde L^\gamma_\gamma \tilde L^\delta_\gamma = \tilde L^\delta_\gamma + \fO(p^{-2})
\ee
we have 
\be
	S^\alpha_\beta(j_\gamma) = \sum_{\delta\not=\gamma}\left( L^\delta_\gamma(j_\gamma) U^{\dag, \alpha}_\delta(j_\gamma) U^\gamma_\beta(j_\gamma)
	+U^{\dag, \alpha}_\gamma(j_\gamma) L^\gamma_\delta(j_\gamma) U^\delta_\beta(j_\gamma)\right)
\ee
\[
	+ p(1-\mu(\gamma))U^{\dag,\alpha}_\gamma(j_\gamma)U^\gamma_\beta(j_\gamma)
	+ p^{-1}\sum_{\delta,\sigma \not=\gamma} L^{\delta}_\gamma(j_\gamma) U^{\dag,\alpha}_{\delta}(j_\gamma) L^\gamma_{\sigma}(j_\gamma) U^{\sigma}_\beta(j_\gamma).
\]
In matrix form, this is
\be
	S^\alpha_\beta(j_\gamma) = pU^\dag\Lambda_\gamma U 
	+ U^\dag \{L,\Lambda_\gamma\} U
	+ p^{-1}U^\dag \fL^\gamma U,
\ee
where
\be
	\fL^\alpha_\beta(j_\gamma) = L^\alpha_\gamma(j_\gamma) L^\gamma_\beta(j_\gamma)
	-p^2 \mu(\gamma)[\Lambda_\gamma]^\alpha_\beta.
\ee 
This proves (\ref{2:1}). With this, we proceed to calculate
\be
	\tr[S(j_\gamma)S(j_\eta)] = \sum_{i=1}^2X^i(\gamma,\eta)
	 + \sum_{i=3}^4(X^i(\gamma,\eta) + X^i(\eta,\gamma)) + \fO(p^{-2}),
\ee
with
\begin{align}
X^1(\gamma,\eta) := & & p^2 \tr[U^\dag(j_\gamma) \Lambda_\gamma U(j_\gamma) U^\dag(j_\eta)\Lambda_\eta U(j_\eta)] \\
X^2(\gamma,\eta) := & & \tr[U^\dag(j_\gamma)\{L(j_\gamma),\Lambda_\gamma\} U(j_\gamma)U^\dag(j_\eta) \{L(j_\eta),\Lambda_\eta\} U(j_\eta)] \\ 
X^3(\gamma,\eta) := & & p\tr [U^\dag(j_\gamma)\{L(j_\gamma), \Lambda_\gamma\} U(j_\gamma)U^\dag(j_\eta) \Lambda_\eta U(j_\eta)] \\
X^4(\gamma,\eta) := & & \tr[U^\dag(j_\gamma)\fL(j_\gamma) U(j_\gamma) U^\dag(j_\eta)\Lambda_\eta U(j_\eta)] \\
\end{align}
Since the matrices $U,L,\fL$ are evaluated at different sites, we Taylor expand. For example,
\be
	U(j_\gamma) = U(nj + (\gamma-1)) = U(j_\eta) + (\eta-\gamma)\partial_x U(j_\eta) + \frac{1}{2}(\eta-\gamma)^2 \partial_x^2 U(j_\eta) + \cdots
\ee
We assume the derivate is uniform ( $\partial_x U(j_\eta) = \partial_x U(j'_\lambda)$), and consider each of the above terms separately. Since $L$ characterizes a fluctuation, we treat it as the same order as $\partial U$. Finally, we suppress the argument $j_\gamma$ of each matrix throughout. Then:
\begin{itemize}
\item Term 1:
\be
	X^1(\gamma,\eta) \approx p^2 \tr[\Lambda_\gamma\Lambda_\eta + (\eta-\gamma)^2(U\partial_x U^\dag \Lambda_\gamma \partial_x UU^\dag \Lambda_\eta - \Lambda_\gamma\Lambda_\eta \partial_x U \partial_x U^\dag)].
\ee
Since $\Lambda_\gamma\Lambda_\eta=0$ for $\gamma\not=\eta$, this simplifies to
\be
	X^1(\gamma,\eta) \approx p^2(\eta-\gamma)^2\tr[ U\partial_x U^\dag\Lambda_\gamma \partial_x U U^\dag \Lambda_\eta].
\ee
\item Term 3:
\be
	X^3(\gamma,\eta) \approx p \tr[\{L,\Lambda_\eta\}\Lambda_\gamma] + (\eta-\gamma)p\tr[ \{L,\Lambda_\eta\} U\partial_x U^\dag \Lambda_\gamma]
	+ (\eta-\gamma) \tr[\{L,\Lambda_\eta\}\Lambda_\gamma\partial_x UU^\dag]
\ee
Since the first term is a product of a diagonal and an off-diagonal matrix, its trace vanishes. What remains is a commutator:
\be
	X^3(\gamma,\eta) = (\eta-\gamma)p\tr\Big[ [\{L,\Lambda_\eta \}, \Lambda_\gamma ] \partial_x U U^\dag\Big],
\ee
which simplifies to
\be
	X^3(\gamma,\eta)= (\eta-\gamma)p\left( L^\gamma_\eta [U\partial_x U^\dag]^\eta_\gamma 
	+ L^\eta_\gamma[\partial_x U U^\dag]^\gamma_\eta\right).
\ee
Note that $X^3(\gamma,\eta) = X^3(\eta,\gamma)$. 
\item Term 4:
Since $\fL$ contains two powers of $L$, we only have to expand $U$ to zeroth order. We find
\be
	X^4(\gamma,\eta) = |L^\gamma_\eta|^2 = X^4(\eta,\gamma).
\ee
\item Term 2:
A similar calculation shows that
\be
	X^2(\gamma,\eta) = 2|L^\gamma_\eta|^2 = X^2(\eta,\gamma).
\ee
\end{itemize}

Finally, combining the results of these five calculations, we find
\be \label{app:result:1}
	\tr[S(j_\gamma) S(j_\eta)] 
	= p^2(\eta - \gamma)^2 \tr  U\partial_x U^\dag \Lambda_\gamma \partial_x U U^\dag \Lambda_\eta
\ee
\[
	+	2(\eta-\gamma)p\left( L^\eta_{\gamma}[\partial_x U U^\dag]^\gamma_{\eta} 
	+ L^\gamma_{\eta}[U \partial_x U^\dag]^\eta_{\gamma}\right)
	+4 |L^\eta_{\gamma}|^2  + \text{const.}
\]
which is (\ref{t1t5}).

\subsection{Berry Phase Term} \label{app:derivation:2}

Using (\ref{qft:l2}), we have
\be
	\partial_\tau \phi^\alpha_\beta  = \sum_\gamma \partial_\tau \tilde L^\alpha_\gamma U^\gamma_\beta + \tilde L^\alpha_\gamma\partial_\tau U^\gamma_\beta,
\ee
where $\tilde L$ is defined below (\ref{qft:l3}). We neglect time derivatives of $\tilde L$, which are already small fluctuations. Then we have
\be
	\vphi^*_\alpha \cdot \partial_\tau \vphi^\alpha
	= \sum_{\delta,\gamma,\beta} \tilde L^\delta_\alpha U^{\dag,\beta}_\delta\tilde L^\alpha_\gamma \partial_\tau U^\gamma_\beta
\ee
\[
	= \sum_{\delta\not=\alpha} \sum_\beta \left[ L^\delta_\alpha U^{\dag,\beta}_\delta \partial_\tau U^\alpha_\beta
	+ U^{\dag,\beta}_\alpha \tilde L^\alpha_\delta \partial_\tau U^\delta_\beta
	+(1-\mu(\alpha)) U^{\dag,\beta}_\alpha \partial_\tau U^\alpha_\beta\right]+ \fO(p^{-2})
\]
\be \label{app:result:2}
= \tr[\Lambda_\alpha \partial_\tau U U^\dag] + p^{-1}\tr[ \{\Lambda_\alpha, L\} \partial_\tau U U^\dag] +\fO(p^{-2}).
\ee
	
\subsection{Integrating out $L$} \label{app:derivation:3}

The Lagrangian terms involving a given matrix element $L^\alpha_\beta$ are:
\be
	  4(J_t + J_{n-t})|L^\alpha_\beta|^2 
	- 2L^\alpha_\beta\left(  [\partial_\tau U U^\dag]^\beta_\alpha  + p((n-t)J_{n-t} - t J_t)[\partial_x U U^\dag]^\beta_\alpha\right)
\ee
\[
	-2L^{\beta}_\alpha \left(  [\partial_\tau U U^\dag]^\alpha_\beta - p((n-t)J_{n-t} - t J_t)[\partial_x U U^\dag]^\alpha_\beta\right)
\]
where $t:= |\alpha-\beta|$. The $\partial_\tau$-dependent terms have come from the Berry term (\ref{topterm}), and the $\partial_x$-dependent terms have come from
\be
	J_t \tr[S(j_\alpha)S(j_\beta)] + J_{n-t} \tr[ S(j_\beta)S(j_{n+\alpha})]
\ee
in the Hamiltonian. Integrating over $L^\alpha_\beta$, and using the identity 
\be
	\int dz dz^* e^{-z^* \omega z + u z + v z^*} = \frac{\pi}{\omega }e^{uv/\omega}
\ee
we are left with a real term,
\begin{equation} \label{intout}
	\fL^{\text{real}}_{\alpha\beta} =   \frac{1}{n(J_t + J_{n-t})}  \tr[ \Lambda_\alpha U \partial_\tau U^\dag \Lambda_\beta \partial_\tau U U^\dag]- p^2 \frac{[(n-t) J_{n-t} - tJ_t]^2}{n(J_t + J_{n-t})} \tr [\Lambda_\alpha U \partial_x U^\dag \Lambda_\beta \partial_x U U^\dag]
\end{equation}
as well as an imaginary term
\be \label{intout2}
	\fL_{\alpha\beta}^{\text{imag}} = p\frac{((n-t)J_{n-t} - tJ_t)}{n(J_t + J_{n-t})}\left( [\partial_x U U^\dag]^\alpha_\beta [\partial_\tau U U^\dag]_\alpha^\beta - [\partial_\tau U U^\dag]^\beta_\alpha [\partial_x U U^\dag]^\alpha_\beta\right).
\ee
The factor of $n$ in the denominator comes from converting the sum over lattice sites with $n$-site unit cell, to an integral. To these terms, we must add the $L$-independent terms appearing in (\ref{app:result:1}) and (\ref{app:result:2}). They modify (\ref{intout}) to 
\be \label{3:33}
	\fL^{\text{real}}_{\alpha\beta} \to  \frac{1}{n(J_t + J_{n-t})}  \tr[ \Lambda_\alpha  U \partial_\tau U^\dag \Lambda_\beta \partial_\tau U U^\dag]
	+p^2 \frac{J_{n-t}J_t n}{(J_t +J_{n-t})}\tr[ \Lambda_\alpha U\partial_x  U^\dag \Lambda_\beta \partial_x U U^\dag].
\ee
Comparing the ratios of the pre-factors of the spatial and imaginary temporal terms, we identify the velocities of the theory as 
\be
	v_t^2  = n^2 p^2 J_{n-t}J_t,
\ee
where $t = |\alpha-\beta|$. This agrees with the flavour wave velocities found in Section \ref{section:fw}. Meanwhile, the terms in (\ref{app:result:2}) modify (\ref{intout2}) to produce the following purely-imaginary contribution to the Lagrangian:

\be \label{3:1}
	\fL^{\text{imag}} =  -\epsilon_{\mu\nu} \sum_{\alpha< \beta} \lambda_{|\alpha-\beta|}   \tr [\partial_\mu U U^\dag \Lambda_\alpha \partial_\nu U U^\dag \Lambda_\beta] - \fS
\ee
where
\begin{equation} \label{sterm}
	\fS := 	\frac{p}{n} \sum_\alpha  \tr[\Lambda_\alpha \partial_\tau U U^\dag]
\end{equation}
and
\be
	\frac{ n \lambda_{t}}{p} := \frac{(n-t)J_{n-t} -tJ_{t}}{J_{t} + J_{n-t}}
\ee
Using the identity $\tr[\partial U U^\dag] = 0$, which is proven in Appendix \ref{app1}, the integral of $\fS$ can be shown to be a total derivative:
\begin{equation}
i\fS = \frac{2\pi p }{n} \sum_{\alpha=2}^n (\alpha-1)Q_\alpha
\end{equation}
where
\begin{equation}
Q_\alpha :=  \frac{1}{2\pi i} \epsilon_{\mu\nu} \int dx d\tau  \tr[\partial_\mu U \partial_\nu U^\dag \Lambda_\alpha].
\end{equation}
Relabelling $\fS = -S_{\text{top}}$, and combining (\ref{3:1}) with (\ref{3:33}), we arrive at (\ref{3:3}).

\section{Proof of $\tr[\partial_\tau UU^\dagger] =0$} \label{app1}

Let $\epsilon^{\alpha_1\alpha_2\cdots \alpha_n}$ be the antisymmetric $n$-tensor, vanishing unless all indices are different in which case it equals $\pm 1$ depending on the sign of the permutation. We can write an arbitrary unitary matrix $U$ as 
\be
 U=\left(\begin{array}{c}\vphi_1\\ \vphi_2 \\ \ldots \\ \vphi_n \end{array}\right)
\ee
where the $\vphi_\alpha$ are orthonormal complex vectors,
\be \label{orth}
\vphi^{\alpha*}\cdot \vphi_\beta=\delta^\alpha_\beta.
\ee
We can write $\vphi_n$ in terms of $\vphi_1, \vphi_2,\ldots \vphi_{n-1}$:
\be \phi_{\alpha_n}^{n*}=\epsilon_{\alpha_1,\alpha_2\ldots \alpha_n}\phi^{\alpha_1}_{1}\phi^{\alpha_2}_{2}\ldots \phi_{n-1}^{\alpha_{n-1}}.
\ee
This follows because
\be \vphi^{n*}\cdot \vphi_1=\epsilon_{\alpha_1,\alpha_2\ldots \alpha_n}\phi^{\alpha_1}_{1}\phi^{\alpha_2}_{2}\ldots \phi_{n-1}^{\alpha_{n-1}}\phi_1^{\alpha_n}
\ee
and 
\be
 \epsilon_{\alpha_1\alpha_2\ldots \alpha_n}\phi_1^{\alpha_1}\phi_1^{\alpha_n}=0.
 \ee
Similarly
\be \vphi^{n*}\cdot \vphi_\alpha=0\ee
for $\alpha=1,2,3, \ldots n-1$.
We use the identity
\be 
\epsilon^{\alpha_1\alpha_2,\ldots \alpha_n}\epsilon_{\beta_1\beta_2\ldots \beta_{n-1}\alpha_n}=\sum_{\{a_1,a_2,\ldots a_{n-1}\}}\hbox{sgn}\{a_1,a_2,\ldots a_{a_{n-1}}\}\delta^{\alpha_1}_{\beta_{a_1}}\delta^{\alpha_2}_{\beta_{a_2}}\ldots 
\delta^{\alpha_{n-1}}_{\beta_{a_{n-1}}}.\label{id}
\ee
Here the sum is over all permutations of $a_1,a_2,\ldots a_{n-1}$. Eq. (\ref{orth}) and (\ref{id}) imply
\be |\vphi_n|^2=1, \ee
\be \vphi_n\cdot \partial \vphi^{n*}=\epsilon^{\alpha_1\alpha_2\ldots \alpha_n}\phi^{1*}_{\alpha_1} \phi^{2*}_{\alpha_2} \ldots \phi^{n-1,*}_{\alpha_{n-1}} \epsilon_{\beta_1\beta_2\ldots \beta_{n-1}\alpha_n}
[(\partial \phi_1^{\beta_1})\phi_2^{\beta_2}\ldots \phi_{n-1}^{\beta_{n-1}}+\ldots ].
\ee
Here the $\ldots$ is a sum over derivatives of each factor. Now we use
\be
 \epsilon^{\alpha_1\alpha_2\ldots \alpha_n}\phi^{1*}_{\alpha_1} \phi^{2*}_{\alpha_2} \ldots \phi^{n-1,*}_{\alpha_{n-1}} \epsilon_{\beta_1\beta_2\ldots \beta_{n-1}\alpha_n}
(\partial \phi_1^{\beta_1})\phi_2{\beta_2}\ldots \phi_{n-1}^{\beta_{n-1}}=\vphi^{1*}\cdot \partial \vphi_1,
\ee
which follows from Eqs. (\ref{orth}) and (\ref{id}). So
\be 
\vphi_n\cdot \partial \vphi^{n*}=\vphi^{1*}\cdot \partial \vphi_1+\vphi^{2*}\cdot \partial \vphi_2+\ldots +\vphi^{n-1,*}\cdot \partial \vphi_{n-1}.
\ee
Thus
\be  \vphi^{n*}\cdot \partial \vphi_{n}=-\vphi^{1*}\cdot \partial \vphi_1-\vphi^{2*}\cdot \partial \vphi_2-\ldots -\vphi^{n-1,*}\cdot \partial \vphi_{n-1},
\ee
so
\be \sum_{\alpha=1}^n\partial \vphi_\alpha\cdot \vphi^{\alpha*}=\hbox{tr}[\partial UU^\dagger] =0.\ee

\section{Factorization of SU($n$) Matrices} \label{app:factorize}
In this appendix, we prove a factorization identity for SU($n$) matrices (\ref{eq:factorize}). Let greek letters index the diagonal generators of SU($n$), lower case roman letters index the off-diagonal generators of SU($n$), and upper case roman letters index the full set of generators. That is,
\be
	\sum_A T_A = \sum_a T_a + \sum_\gamma T_\gamma.
\ee
Then, given $U = e^{i\theta_AT_A} \in \mbox{SU}(n)$, we may factorize it as follows:
\be
	U = e^{i\phi_\gamma T_\gamma}e^{i\phi_aT_a}.
\ee

We will prove this identity to third order in the $\phi$ and $\theta$, but mention how the proof extends to every order in perturbation theory.

\emph{Proof:}
Using the Baker-Campbell-Hausdorff formula, 
\be \label{eq:factor:1}
	\log(e^Xe^Y) = X + Y + \frac{1}{2}[X,Y] + \frac{1}{12}\left( [X,[X,Y]] -[Y,[X,Y]]\right) + \cdots
\ee
we have
\be \label{eq:factor:2}
	\log e^{i\phi_\gamma T_\gamma}e^{i\phi_aT_a}
	= i\phi_AT_A - \frac{1}{2}\phi_\gamma \phi_a[T_\gamma,T_a]
	-\frac{i}{12}\left( \phi_\gamma \phi_\beta \phi_a [T_\gamma, [T_\beta ,T_a]]
	- \phi_\gamma \phi_a \phi_b [T_a, [T_\gamma, T_b]\right)
	+\fO(\phi^4)
\ee
which equals 
\be \label{eq:factor:3}
	=i\Big[ \phi_A- \phi_\gamma \phi_bf_{\gamma bA}
	+\frac{1}{3}\left( \phi_\gamma \phi_\beta \phi_b f_{\beta b C}f_{\gamma CA}
	- \phi_\gamma \phi_d \phi_bf_{\gamma bC} f_{dCA}\right)\Big]T_A
	+\fO(\phi^4).
\ee

The formula for the higher order terms occurring (\ref{eq:factor:1}) and (\ref{eq:factor:2}) are quite complicated, but always involve nested commutators. This important fact allows us to reduce every term in the expansion to one that is linear in the generators, $T_A$. A term that is $\sim \phi^n$  will involve $n-1$ nested commutators, leading to a contribution that is proportional to a product of $n-1$ structure factors $f_{abc}$, multiplied by a single SU($n$) generator $T_A$.  Therefore, order-by-order, we may construct a mapping between the $\theta_A$ and the $\phi_A$:
\be
	\theta_A = \phi_A- \phi_\gamma \phi_bf_{\gamma bA}
	+\frac{1}{3}\left( \phi_\gamma \phi_\beta \phi_b f_{\beta b C}f_{\gamma CA}
	- \phi_\gamma \phi_d \phi_bf_{\gamma bC} f_{dCA}\right) + \fO(\phi^4).
\ee

To prove the factorization identity, we must be able to invert this formula. This is done by a repeated application of 
\be
	\phi_A = \theta_A + \phi_\gamma \phi_bf_{\gamma bA} -\frac{1}{3}\left( \phi_\gamma \phi_\beta \phi_b f_{\beta b C}f_{\gamma CA}
	- \phi_\gamma \phi_d \phi_bf_{\gamma bC} f_{dCA}\right)  + \fO(\phi^4).
\ee
into each of the terms on the RHS. We find:
\be
	\phi_A = \theta_A + \theta_\gamma \theta_b f_{\gamma bA}
	+ \frac{2}{3}\theta_\gamma  \theta_\beta \theta_b f_{\beta be}f_{\gamma eA}
	+ \frac{1}{3}
	\theta_\gamma \theta_d \theta_bf_{\gamma bC} f_{dCA} + \fO(\theta^4)
\ee
Thus, for any SU($n$) matrix $U=e^{i\theta_AT_A}$, we may perform this transformation to obtain the factorized form occurring above.$\square$

\section{Goldstone Mode Expansion of the Action} \label{app:goldstone}

In this appendix, we derive (\ref{eq:gold1}). We use lower case roman letters to index the off-diagonal generators, and upper case latin letters to index the complete set of generators. We start with 
\be
	\partial_\mu U U^\dag 
	= i\partial_\mu \theta_aT_a + \left[\partial_\mu \theta_a \theta_b  -\frac{1}{2} \partial_\mu(\theta_a\theta_b)\right]T_aT_b
	-\frac{i}{2}\left[ \partial_\mu\theta_a \theta_b\theta_c - \partial_\mu(\theta_a\theta_b)\theta_c
	+ \frac{1}{3} \partial_\mu(\theta_a\theta_b\theta_c)\right] T_a T_bT_c 
	+ \fO(\theta^4).
\ee

Since
\be
	\partial_\mu\theta_a\theta_b - \frac{1}{2}\partial_\mu(\theta_a\theta_b)
	= \frac{1}{2}\left[\partial_\mu\theta_a\theta_b  -\theta_a\partial_\mu \theta_b\right],
\ee
we have 
\be
	 \left[\partial_\mu \theta_a \theta_b  -\frac{1}{2} \partial_\mu(\theta_a\theta_b)\right]T_aT_b
	 = \frac{1}{2}\partial_\mu\theta_a\theta_b[T_a,T_b]
	 =i\partial_\mu \theta_a\theta_b  f_{abC}T_C.
\ee

Since
\be
	\partial_\mu\theta_a\theta_b\theta_c - \partial_\mu(\theta_a\theta_b)\theta_c
	+\frac{1}{3}\partial_\mu(\theta_a\theta_b\theta_c)
	=\frac{1}{3}( \partial_\mu\theta_a\theta_b\theta_c  - \theta_a\partial_\mu\theta_b\theta_c)
	+\frac{1}{3}(\theta_a\theta_b\partial_\mu\theta_c - \theta_a\partial_\mu\theta_b\theta_c),
\ee
we have
\be
\left[ \partial_\mu\theta_a \theta_b\theta_c - \partial_\mu(\theta_a\theta_b)\theta_c
	+ \frac{1}{3} \partial_\mu(\theta_a\theta_b\theta_c)\right] T_a T_bT_c 
\ee
\[
	=- \frac{1}{3}\theta_a\partial_\mu \theta_b\theta_c\left( [T_a,T_b]T_c +T_a[T_b,T_c]\right)
	= -\frac{2i}{3}\theta_a\partial_\mu\theta_b\theta_c f_{abD}[T_D,T_c]
	= \frac{4}{3}\theta_a\partial_\mu\theta_b\theta_c f_{abD}f_{DcE}T_E.
\]

Therefore, we have
\be
	\partial_\mu U U^\dag = i\partial_\mu \theta_a T_a + i\partial_\mu\theta_a\theta_b f_{abC}T_C
	-\frac{2i}{3}f_{abD}f_{DcE}\theta_a\partial_\mu\theta_b\theta_c T_E	+\fO(\theta^4).
\ee

This yields 
\be
	-\tr[ \partial_\mu U U^\dag \Lambda_\alpha\partial_\mu U U^\dag \Lambda_\beta]
	=  \partial_\mu\theta_a\partial_\mu\theta_b\tr [T_a\Lambda_\alpha T_b\Lambda_\beta]
\ee
\[
	+  \left[f_{bcE} +\frac{2}{3}f_{bcD}f_{DgE}\theta_g\right]\partial_\mu\theta_a\partial_\mu\theta_b\theta_c\left( \tr[T_a\Lambda_\alpha T_E\Lambda_\beta] +  \tr[T_E\Lambda_\beta T_a\Lambda_\alpha]\right)
\]
\[
	+\partial_\mu\theta_a\theta_b\partial_\mu\theta_c\theta_d f_{abE}f_{cdG} \tr\Lambda_\alpha T_E\Lambda_\beta T_G.
\]
Now we want to simplify this by understanding
\be
	\tr[T_a \Lambda_\alpha T_b \Lambda_\beta]
	=[T_a]_{\beta\alpha} [T_b]_{\alpha\beta}
\ee

Since $\alpha\not=\beta$, $\tr[ T_a \Lambda_\alpha T_b \Lambda_\beta]$ vanishes if either of $a$ or $b$ is a diagonal generator. All of the off-diagonal generators have the same structure in SU($n$) (discussed in the main text). Using the notation introduced above, we have 
\be
	\tr[T_a \Lambda_\alpha T_b\Lambda_\beta]
	+\tr[T_b \Lambda_\alpha T_a\Lambda_\beta]  = \begin{cases}2 \delta_{ab}  & a,b \in I_{\alpha\beta}\\
	0 & \text{else} \\
	\end{cases}.
\ee
Returning to our calculation, we now have 
\be
	-\tr[\partial_\mu U U^\dag \Lambda_\alpha \partial_\mu U U^\dag \Lambda_\beta]
	=\sum_{a\in I_{\alpha\beta} }\Big[ (\partial_\mu \theta_{a}^2)    + 2f_{bca}\partial_\mu\theta_a\partial_\mu\theta_b\theta_c
	+
	 \frac{4}{3}f_{bcE}f_{Eda}\partial_\mu\theta_{a}\partial_\mu\theta_b\theta_c\theta_d 
	+\partial_\mu\theta_e\theta_b\partial_\mu\theta_c\theta_d f_{eba}f_{cda} \Big]
\ee

where all repeated indices are summed over.

\section{Renormalization Group Calculations} \label{app:rgcalc}

\begin{figure}[h]
\includegraphics[width = .3\textwidth]{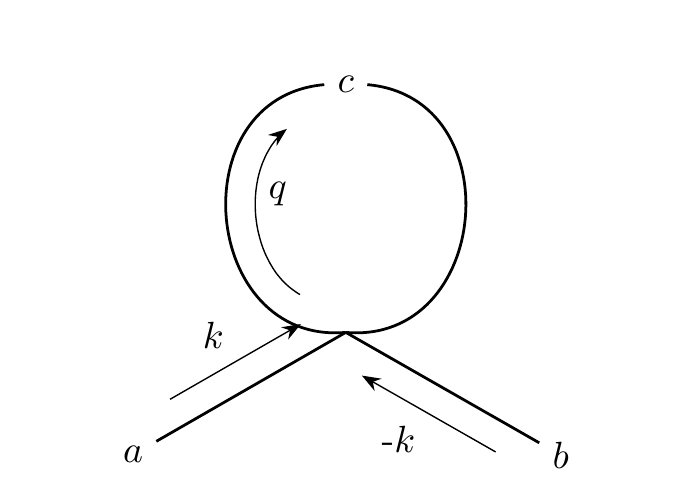}
\caption{The diagram $\Pi_{ab}(k)$, drawn using [\onlinecite{ellis2017tikz}].}
\label{fig:feynman}
\end{figure}

We use dimensional regularization to evaluate one-loop diagrams in $d = 2-\epsilon$ dimensions in (\ref{eq:2:renorm}). We drop all `$r$' superscripts, and introduce the following compact notation:
\be \label{eq:gg:1}
	g_{abcd}^{(1)}(\mu) := \frac{M^\epsilon}{4}\sqrt{g_ag_bg_cg_d} h_e(\mu) \frac{f_{ace}f_{bde}}{g_e}
\ee
\be \label{eq:gg:2}
	g_{abcd}^{(2)}(\mu) := \frac{M^\epsilon h_a(\mu) }{3}\frac{\sqrt{g_ag_bg_cg_d}}{g_a} f_{bcE}f_{Eda}
\ee 
Again, all indices refer to off-diagonal SU($n$) generators, except for the upper case letters, which refer to the complete set. We've introduced a renormalization scale $M$ so that the coupling constants remain dimensionless. Since we are only tasked with calculating the $\{Z^\mu_a\}$, the only diverging diagrams we must consider are those that correct the boson self energy. This immediately implies that the cubic interaction term occurring in (\ref{eq:2:renorm}) plays no effect at this order. The only contributing diagram $\Pi_{ab}(k)$, shown in Figure \ref{fig:feynman}, equals 
\be \label{pro:1}
	\Pi_{ab}(k) = -2\int \frac{d^d q}{(2\pi)^d} \langle \theta_c (q)\theta_c(-q)\rangle \left[ g_{abcc}(\mu) k_\mu k_\mu + g_{ccab}(\mu) q_\mu q_\mu\right]
\ee
where
\be
	g_{abcd} = g^{(1)}_{abcd} + g^{(2)}_{abcd}.
\ee 
In addition to UV divergences, there are also IR divergences occurring at zero momenta. To remove these, we introduce a small mass $m$ to the boson fields $\theta$, and take the limit $m\to 0$ once we've extracted the UV divergence. A convenient mass term with the appropriate dimensions is $m^2\bar v^3u_a \theta^2$. 
Then, the free propagator is 
\be
	\langle \theta_c (q)\theta_c (-q)\rangle =  \frac{u_c}{\omega^2 +u_c^2 \vec{q}^2 +  m^2u_c^2\bar v^3} \hspace{10mm} q = (\omega ,\vec{q})
\ee
and we have two integrals to consider:
\subsubsection{Two Integrals:} \label{int}

\begin{itemize}
\item{}\textbf{Integral 1:} 
\be
	\int \frac{d^dq}{(2\pi)^d} \langle \theta_c (q) \theta_c(-q)\rangle
	= \frac{1}{2}  \int \frac{d^{d-1} q}{(2\pi)^{d-1}}\frac{1}{\sqrt{\vec{q}^2 +m^2  \bar v^3}}
		=  \frac{1}{2\pi \epsilon  } + \fO(\epsilon^0)
\ee
\item{} \textbf{Integral 2:}
\be
	\int \frac{d^d q}{(2\pi)^d} \langle \theta_c(q)\theta_c(-q)\rangle q_x q_x = \frac{1}{2}  \int \frac{d^{d-1} q}{(2\pi)^{d-1}}\frac{\vec{q}^2}{\sqrt{\vec{q}^2 +m^2 \bar v^3 }}
	= \frac{1}{2\pi \epsilon  } m^2  \bar v^3  + \fO(\epsilon^0)
\ee
where we've taken $\mu=x$ without loss of generality. It appears that such integrals will renormalize the boson masses; however, since these contributions are proportional to the IR cutoff $m$, when we restore $m \to 0$, these poles will drop out of our calculations. See equation 13.82 of [\onlinecite{Peskin:1995ev}] for a similar argument in the O(3) nonlinear sigma model. 
\end{itemize}

Returning to the process (\ref{pro:1}), we find that 
\be \label{result1}
	\Pi_{ab}(k) = -\frac{ 1 }{\pi \epsilon} k_\mu k_\mu   g_{abcc}(\mu)
	= -\frac{M^\epsilon \sqrt{g_ag_b}}{\pi\epsilon} k_\mu k_\mu g_c\left[ \frac{1}{4}h_e(\mu) \frac{f_{ace}f_{bce}}{g_e}
	- \frac{h_a(\mu)}{3g_a}  f_{bcE}f_{acE}\right].
\ee
This result will contribute to the renormalization constants involving $\theta_a$ and $\theta_b$. In the following subsections, we will use properties of the SU($n$) structure factors, $f_{abc}$, to simplify both of the terms occurring in (\ref{result1}).

\subsection{Lemma 1}

Here we prove 
\be \label{lemma1}
	\frac{g_ch_e(\mu) }{g_e} f_{ace}f_{bce} = \delta_{ab} \sum_{\substack{i=1 \\ i\not=t}}^{n-1} \frac{h_i(\mu)}{g_i}g_{|t-i|}
\ee
where $t:= |\alpha-\beta|\Big|_{I_{\alpha\beta} \ni a}$ and $g_x := g_{x \text{ mod } n}$ for $x>n$.

\vspace{5mm}

\emph{Proof:}
Since $a,c,e$ all correspond to off-diagonal generators, $f_{ace}$ will vanish unless 
\be
	I_a \not= I_c \not= I_e \not= I_a.
\ee
Moreover, for $a$ and $e$ fixed, there is a unique value of $c$ such that $f_{ace}\not=0$. Calling this value $c_*$, we then have
\be
	 \frac{g_c h_e(\mu)}{g_e} f_{ace}f_{bce}
	= \frac{1}{4}\delta_{ab}\frac{g_{c_*}h_e(\mu)}{g_e} \hspace{5mm} \text{(no sum over $e$) }
\ee
since $f_{bc_*e}=0$ unless $a=b$, and all purely off-diagonal structure factors in SU($n$) have magnitude $\frac{1}{2}$. Moreover, one can verify explicitly that for $a\in I_{\alpha\beta}$ and $e\in I_{\gamma\delta}$, with $I_{\alpha\beta} \cap I_{\gamma\delta}  = \emptyset$, 
\be
	g_{c_*} = \delta_{\alpha\gamma}g_{|\beta-\delta|}
	+\delta_{\alpha\delta} g_{|\beta-\gamma|}
	+\delta_{\beta\gamma} g_{|\alpha-\delta|}
	+\delta_{\beta\delta} g_{|\alpha-\gamma|}.
\ee
(Note that if $\{\alpha,\beta\} \cap = \{\gamma,\delta\} = \emptyset$, $[T_a,T_e]=0$.) Therefore, writing $\sum_e = \sum_{\gamma <\delta} \sum_{e \in I_{\gamma\delta}}$, the left hand side of (\ref{lemma1}) is
\be
	\frac{g_c h_e(\mu)}{g_e}f_{ace}f_{bce}
	=\frac{1}{4}\delta_{ab} \sum_{\gamma <\delta}^n \frac{h_e(\mu)}{g_e} \sum_{\substack{e\in I_{\gamma\delta}\\
	e\not\in I_{\alpha\beta}} }\left[ \delta_{\alpha\gamma}g_{|\beta-\delta|}
	+\delta_{\alpha\delta} g_{|\beta-\gamma|}
	+\delta_{\beta\gamma} g_{|\alpha-\delta|}
	+\delta_{\beta\delta} g_{|\alpha-\gamma|}\right] 
\ee
\[
	=\frac{\delta_{ab}}{2} \sum_{\substack{\gamma <\delta \\ I_{\gamma\delta} \not= I_{\alpha\beta} }}^n \frac{h_{|\delta-\gamma|}(\mu)}{g_{|\delta-\gamma|}} \left[ \delta_{\alpha\gamma}g_{|\beta-\delta|}
	+\delta_{\alpha\delta} g_{|\beta-\gamma|}
	+\delta_{\beta\gamma} g_{|\alpha-\delta|}
	+\delta_{\beta\delta} g_{|\alpha-\gamma|}\right].
\]
We simplify each of these four terms. Let $t := \beta-\alpha >0$ (we assume without loss of generality that $\alpha <\beta$). Then:
\begin{itemize}
\item
\be
 \sum_{\substack{\gamma <\delta \\ I_{\gamma\delta} \not= I_{\alpha\beta} }}^n \frac{h_{|\delta-\gamma|}(\mu)}{g_{|\delta-\gamma|}} \delta_{\alpha\gamma}g_{|\beta-\delta|}
=\sum_{\substack{\delta=\alpha+1\\
\delta \not=\beta}}^n \frac{h_{|\delta-\alpha|}(\mu)}{g_{|\delta-\alpha|}} g_{|\beta-\delta|}
=\sum_{\substack{i=1 \\ i\not=t}}^{n-\alpha} \frac{h_i(\mu)}{g_i} g_{|t-i|}
\ee
\item
\be \label{eq:sum:2}
	\sum_{\substack{\gamma <\delta \\ I_{\gamma\delta} \not= I_{\alpha\beta} }}^n \frac{h_{|\delta-\gamma|}(\mu)}{g_{|\delta-\gamma|}} 
	\delta_{\alpha\delta} g_{|\beta-\gamma|}
	=\sum_{\gamma=1}^{\alpha-1}\frac{ h_{|\alpha-\gamma|}(\mu)}{g_{|\alpha-\gamma|}}g_{|\beta-\gamma|}
	=\sum_{i=n-\alpha+1}^{n-1} \frac{h_i(\mu)}{g_i(\mu)} g_{|i-t|}
\ee
\item
\be
	\sum_{\substack{\gamma <\delta \\ I_{\gamma\delta} \not= I_{\alpha\beta} }}^n \frac{h_{|\delta-\gamma|}(\mu)}{g_{|\delta-\gamma|}} 
	\delta_{\beta\gamma} g_{|\alpha-\delta|}
 = \sum_{\delta=\beta+1}^n\frac{h_{|\delta - \beta|}(\mu)}{g_{|\delta-\beta|}}g_{|\alpha-\delta|}
 =\sum_{i=1}^{n-\beta} \frac{h_i(\mu)}{g_i} g_{|t+i|}
\ee
\item 
\be \label{eq:sum:4}
	\sum_{\substack{\gamma <\delta \\ I_{\gamma\delta} \not= I_{\alpha\beta} }}^n \frac{h_{|\delta-\gamma|}(\mu)}{g_{|\delta-\gamma|}}\delta_{\beta\delta} g_{|\alpha-\gamma|}
	= 
	\sum_{\substack{\gamma=1 \\
	\gamma\not=\alpha}}^{\beta-1} \frac{h_{|\beta-\gamma|}(\mu)}{g_{|\beta-\gamma|}}g_{|\alpha-\gamma|}
	= \sum_{\substack{i=n-\beta+1 \\
	i\not= n-t}}^{n-1} \frac{h_i(\mu)}{g_i} g_{|i+t|}
\ee
 \end{itemize} 
where it is understood that $g_x := g_{x \text{ mod } n}$ for $x >n$. In (\ref{eq:sum:2}) and (\ref{eq:sum:4}), we used the fact that $g_i = g_{n-i}$ and $h_i = h_{n-i}$ in the last equations. Combining these results, we have
\be
	 \frac{g_c h_e(\mu)}{g_e}f_{ace}f_{bce}
	= \frac{1}{2}\delta_{ab} \left[ \sum_{\substack{i=1 \\ i\not=t}}^{n-1} \frac{h_i(\mu)}{g_i}g_{|t-i|}
	+ \sum_{\substack{i=1 \\ i\not=n-t}}^{n-1} \frac{h_i(\mu)}{g_i}g_{|t+i|}\right].
\ee
Finally, replacing $i \to n-i$ in the second sum, we see that these two terms are in fact. Therefore, we arrive at 
\be
	\frac{g_c h_e(\mu)}{g_e}f_{ace}f_{bce}
	= \delta_{ab} \sum_{\substack{i=1 \\ i\not=t}}^{n-1} \frac{h_i(\mu)}{g_i}g_{|t-i|},
\ee
which completes the proof.$\square$

\subsection{Lemma 2}

Here we prove 
\be
	g_c f_{bcE}f_{acE} = \frac{1}{2}\delta_{ab}\left[ g_a + \frac{1}{2}\sum_c g_c\right]
\ee

\emph{Proof:}

We first write
\be
	 g_c f_{bcE}f_{acE} =\sum_{\gamma<\delta}^n \sum_{c\in I_{\gamma\delta}} g_c f_{bcE}f_{acE}.
\ee
If $c\in I_a$, then $f_{bcE}f_{acE}$ vanishes unless $b=a$, and in this case equals  
\be
	\delta_{ab}\sum_E \left[ f_{a \bar a E}\right]^2 = \delta_{ab}, \hspace{10mm}
\ee
where $\bar a$ is the unique index satisfying $\bar a\in I_a$ with $\bar a \not=a$. Indeed, for $a,\bar a \in I_{\alpha\beta}$, we have
\be
	[T_a, \bar T_a] = \pm 2i (\Lambda_\alpha - \Lambda_\beta).
\ee
Since $\Lambda_\alpha - \Lambda_\beta$ generate the traceless diagonal Hermitian matrices, we may take choose them as the diagonal SU($n$) generators. In this case, $f_{a\bar a E} =0$ unless $E$ corresponds to $(\Lambda_\alpha - \Lambda_\beta)$, where it equals 1. Now, if $c\not\in I_a$, then $f_{acE}$ will vanish except for a unique value $e_*$, with $e_*\not\in I_a\cup I_c$. The term $f_{bce}$ forces $a=b$, too. Since $|f_{abc}| = \frac{1}{2}$ for purely off-diagonal generators, we have
\be
	 g_c f_{bcE}f_{acE}
	= \delta_{ab}g_a +\frac{1}{4}\delta_{ab} \sum_{c\not\in I_a} g_c.
\ee
Finally, noting that
\be
	\frac{1}{2}g_a + \frac{1}{4}\sum_{c\not\in I_a} g_c =\frac{1}{4}\sum_c g_c
\ee completes the proof.$\square$

\subsection{Result}

Combining the results of both Lemmas, we conclude that (\ref{result1}) equals 
\be \label{result:post}
	\Pi_{ab}(k) = -\frac{M^\epsilon g_a\delta_{ab}}{2\pi\epsilon} k_\mu k_\mu \Big(\sum_{\substack{i=1 \\ i\not=t}}^{n-1}  \frac{h_i(\mu)}{2g_i}g_{|t-i|}
	- \frac{h_a(\mu)}{3g_a}\big[ g_a + \frac{1}{2}\sum_c' g_c\big]\Big).
\ee
(no sum over $a$). Since 
\be
	 \partial_\mu \theta \partial_\mu \theta \sim
	- \theta \partial_\mu^2 \theta 
	\sim  +  k_\mu k_\mu \theta(k) \theta(-k)
\ee
we may read off from $\Pi_{ab}(k)$ the renormalization group constants:
\be \label{ztau}
Z_a^\tau  = 1 + \frac{M^\epsilon g_a u_a }{2\pi\epsilon} \Big(\sum_{\substack{i=1 \\ i\not=t}}^{n-1}  \frac{1}{u_i g_i}g_{|t-i|}
	- \frac{2}{3g_au_a }\big[ g_a + \frac{1}{2}\sum_c g_c\big]\Big)
\ee
\be \label{zx}
Z_a^x = 1 + \frac{M^\epsilon g_a }{2\pi u_a \epsilon} \Big(\sum_{\substack{i=1 \\ i\not=t}}^{n-1}  \frac{u_i}{ g_i}g_{|t-i|}
	- \frac{2u_a}{3g_a }\big[ g_a + \frac{1}{2}\sum_c g_c\big]\Big)
\ee
(no sum over $a$).

\section{Numerical Verification} \label{app:beta:diff}

In this appendix, we find the beta functions for the velocity differences, $\Delta_{tt'}$, and consider special cases. Assuming the velocities $\{u_t\}$ are initially close together, we rewrite (\ref{beta:result}) to linear order in $\Delta^t$ as
\begin{itemize} 
\item 
\be \label{betaeven}
	\beta_{u_t}^{n=2q} = \frac{g_t}{2\pi} \Big[ \sum_{i=1}^{q-1}\frac{\Delta_{ti}}{g_i}\left( g_{i+t} + g_{|t-i|} \right)
	+ \frac{g_{|t- q|}}{g_q }\Delta_{tq}\Big] + \fO(\Delta^2)
\ee 
\item 
\be \label{betaodd}
	\beta_{u_t}^{n=2q+1} =\frac{g_t}{2\pi} \sum_{\substack{i=1}}^q \frac{\Delta_{ti}}{g_i}\left( g_{i+t} + g_{|i-t|}\right) + \fO(\Delta^2)
\ee 
\end{itemize}
depending on the parity of $n$. (We've introduced a $g_0 :=0$ for notational convenience). Here we have used the fact that only $q := \lfloor \frac{n}{2} \rfloor$ velocities and coupling constants are unique. The beta function for a component $\Delta^i$ of $\bDelta$ (defined in (\ref{eq:vec:del})) is then
\begin{itemize} 
\item 
\be 
	\beta_{\Delta^t}^{n=2q} = \frac{1}{2\pi} \sum_{i=1}^{q-1}\frac{\Delta^i}{g_i}\Big[ g_1
	\left( g_{i+1} + g_{|1-i|} \right)
	- g_t\left( g_{i+t} + g_{|t-i|} \right)\Big]
\ee
\[
		+\frac{g_t\Delta^t}{2\pi}  \Big[\sum_{i=1}^{q-1}\frac{1 }{g_i}\left( g_{i+t} + g_{|t-i|} \right)
		+
	 \frac{g_{|t- q|}}{g_q }\Big]
	  +\frac{\Delta^q}{2\pi g_q} \left[ -g_tg_{|t-q|} + g_1 g_{|1-q|}\right]
 + \fO(\Delta^2)
\]
\item 
\be 
	\beta_{\Delta^t}^{n=2q+1} =\frac{1}{2\pi} \sum_{\substack{i=1}}^q \frac{\Delta^i}{g_i}\Big[ g_1\left( g_{i+1} + g_{|i-1|}\right) 
	-g_t\left( g_{i+t} + g_{|i-t|}\right)\Big]
	+ \frac{g_t\Delta^t}{2\pi} \sum_{\substack{i=1}}^q \frac{1}{g_i}\left( g_{i+t} + g_{|i-t|}\right) 
	 + \fO(\Delta^2).
\ee 
\end{itemize}
depending on the parity of $n$. Clearly, finding the eigenvalues of the $R$ matrix in (\ref{beta:matrix}) is a difficult task. As a first check, we consider the symmetric point where all couplings equal the same value, $g$ (except for the artificial $g_0$, which is always zero). In this case, we can clearly read off from (\ref{beta:result}) that
\be
	\beta_{\Delta^t} = \frac{g\Delta^t}{2\pi}(n-1)
\ee
so that the matrix beta equation is diagonal, with positive eigenvalues. Next, we consider small values of $n$. \begin{itemize}

\item SU(4)

 In this case, there is a single velocity difference, $\beta_{\Delta_{12}}$, with

\be
	\beta_{\Delta_{12}} =  \frac{\Delta_{12}}{2\pi g_2 }\left[ g_1^2 + 2g_2^2\right] >0 
\ee

\item SU(5)

In this case, there is again a single velocity difference, with 
\be
	\beta_{\Delta_{12}} =  \frac{1 }{\pi  g_2 }\Delta_{12} ( g_1^2 + g_2^2) > 0
\ee

\item SU(6)

In this case, there are three velocities, three coupling constants, and two unique velocity differences, $\Delta_{12}$ and $\Delta_{13}$. The eigenvalues of the 2x2 $R$ matrix are
\be
	\text{spec}(R) = \{ \frac{1}{2\pi g_1g_2g_3}\left( g_1^2g_2^2 + g_1^3 g_3 + g_1g_2^2g_3 + g_1^2g_3^2 + g_2^2g_3^2\right), \frac{1}{2\pi g_1g_2g_3}\left( g_1^2 g_2^2 + 2g_1^2g_3^2 + 2g_2^2g_3^2\right) \}
\ee
both of which are positive. 

\end{itemize}

Unable to find the eigenvalues of the $R$ matrix explicitly, we resort to a numerical investigation of its spectrum. We verify that the spectrum is positive definite by computing the minimal eigenvalue of $R$ for fixed coupling constants. First, we choose the $\lfloor \frac{n}{2} \rfloor$ coupling constants randomly from the interval $(0,1)$. In 10 000 trials, we find that the minimal eigenvalue is always strictly positive, for SU($n$) with $n=3,4,\cdots, 50$. Next, we probe points in parameter space where different coupling constants have a common value, by choosing coupling constants from a discrete lattice on $(0,1)^{\lfloor \frac{n}{2} \rfloor}$. Since the dimension of the lattice increases with $n$, we choose a coarser discretization as $n$ increases, to keep the number of lattice points below 100 000. In this case, we find that for $n=3,4,\cdots, 16$, the minimal eigenvalue of the $R$ matrix is again strictly positive. This supports the conjecture that the spectrum of $R$ is always positive, so that each velocity difference $\Delta_{tt'}$ flows to zero in the IR. 


\section{Strong Coupling Analysis} \label{app:strong}

\begin{figure}[htbp]
\begin{center}
\includegraphics[width=0.75\textwidth]{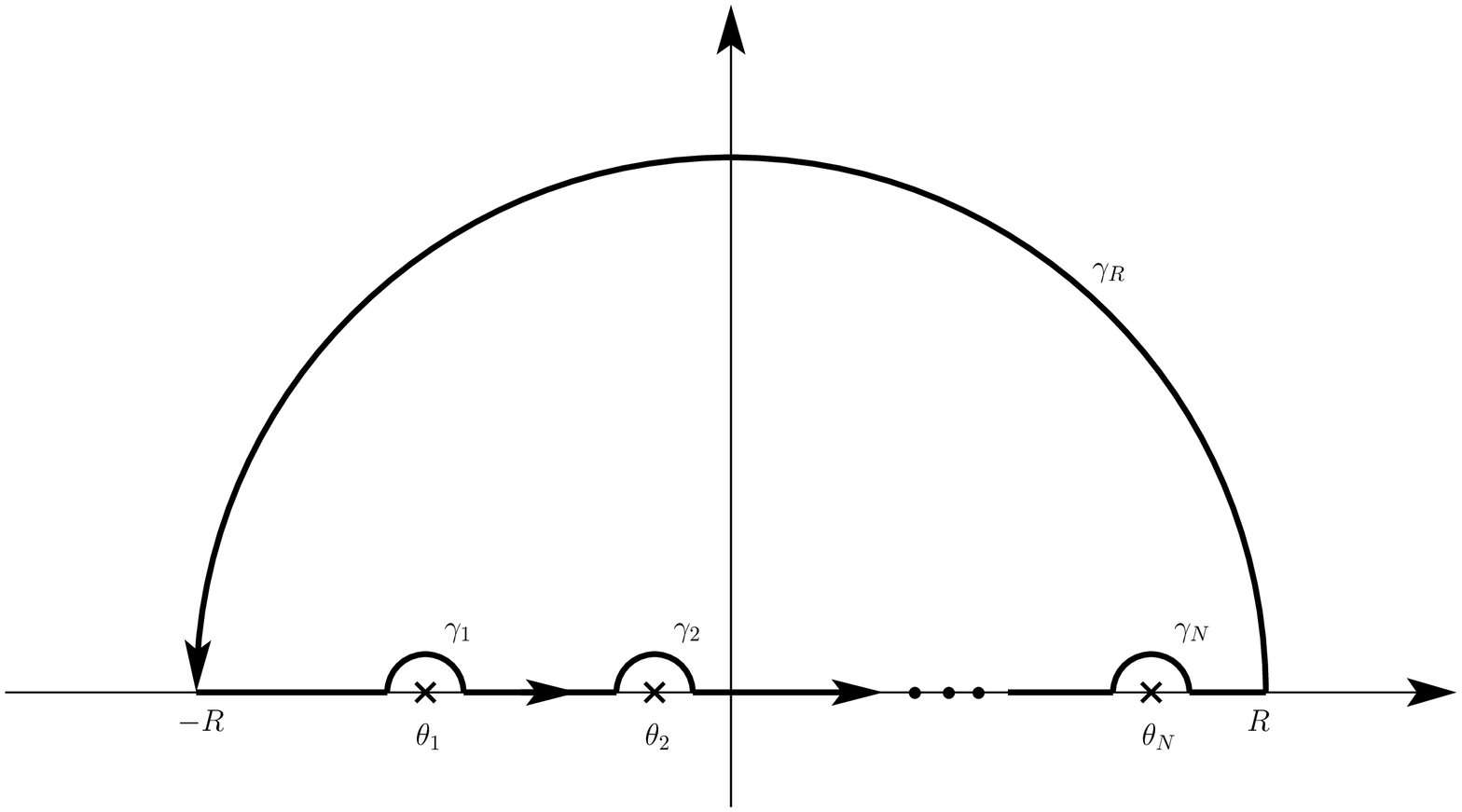}
\caption{Contour integral for evaluating  Eq.~\eqref{eq:Nintegral} for $C(\vec s)\geq 0$. The $\gamma_1,\gamma_2 \dots$ segments are semi-circles of radius $\epsilon$ around $\theta_1,\theta_2 \dots$}
\label{fig:bigsinxxcontour}
\end{center}
\end{figure}

\begin{figure}[htbp]
\begin{center}
\includegraphics[width=0.75\textwidth]{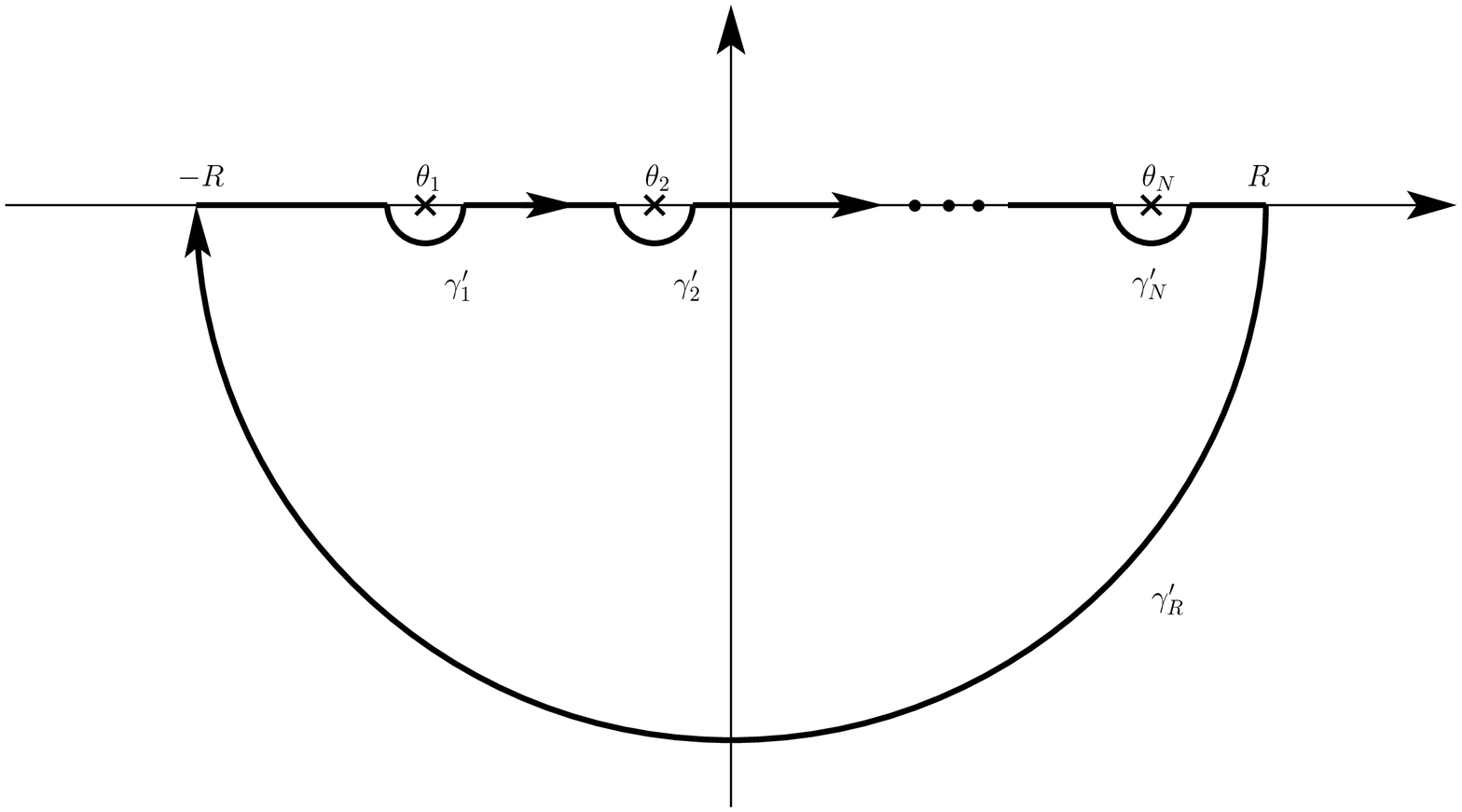}
\caption{Contour integral for evaluating  Eq.~\eqref{eq:Nintegral} for $C(\vec s)\leq 0$. The $\gamma_1,\gamma_2 \dots$ segments are semi-circles of radius $\epsilon$ around $\theta_1,\theta_2 \dots$}
\label{fig:contour2}
\end{center}
\end{figure}

In this appendix, starting with 
\be
	z(\theta_1,\theta_2,\cdots, \theta_n) = \int_{-\infty}^{\infty} \frac{dk}{2\pi} \prod_{\alpha=1}^n \frac{2\sin \left( \frac{1}{2}(k-\theta_\alpha)\right)}{(k-\theta_\alpha)},
\ee
we prove (\ref{st:result1}) and (\ref{st:result2}). We assume that the topological angles are ordered and all different, i.e.\ $\theta_1< \theta_2 <\dots<\theta_n$. First, we split each $\sin$ into parts:
\begin{equation}
\begin{split}
\int \limits_{-\infty}^{\infty} \frac{dk}{2\pi}   \prod \limits_{\alpha=1}^n  \frac{2\sin \big(\frac{1}{2} (k-\theta_\alpha)\big)}{(k-\theta_\alpha)} 
&= \int \limits_{-\infty}^{\infty} \frac{dk}{2\pi}   \prod \limits_{\alpha=1}^n \frac{ \sum \limits_{s_\alpha \in \{-1,+1\}} s_\alpha \exp\big(\frac{i}{2}s_\alpha (k-\theta_\alpha)\big)}{i(k-\theta_\alpha)} \\
&=  \int \limits_{-\infty}^{\infty} \frac{dk}{2\pi}  \frac{1}{i^n\prod\limits_{\alpha=1}^n (k-\theta_\alpha)} \sum_{s_1 \in \{-1,1\}} \sum_{s_2 \in \{-1,1\}}  \dots \sum_{s_n \in \{-1,1\}} \prod\limits_\alpha\Big( s_\alpha\exp\big( \frac{i}{2} s_\alpha( k  - \theta_\alpha) \big)\Big)\\
&=\sum_{\vec s \in \otimes^n\{-1,+1\}}\left[  \exp\Big( -\frac{i}{2}  \sum\limits_{\alpha=1}^n s_\alpha \theta_\alpha \Big)\Big( \prod\limits_\alpha s_\alpha\Big) \int \limits_{-\infty}^{\infty} \frac{dk}{2\pi}  \frac{ \exp\left( \frac{1}{2} i k \sum\limits_{\alpha=1}^n s_\alpha \right)}{i^n\prod\limits_{\alpha=1}^n (k-\theta_\alpha)} \right].
\end{split}
\end{equation}
In every element of this large sum we now have to evaluate an integral of the form
\begin{equation}
\label{eq:Nintegral}
\begin{split}
\int \limits_{-\infty}^{\infty} \frac{dk}{2\pi}  \frac{ \exp\left( \frac{1}{2} i C(\vec s) k  \right)}{\prod\limits_{\alpha=1}^n (k-\theta_\alpha)} 
\end{split}
\end{equation}
where $C(\vec s)= \sum\limits_\alpha s_\alpha$. Each of these terms we can calculated using complex analysis. For instance, if $C(\vec{s}) \geq 0$, we use the contour shown in Fig.~\ref{fig:bigsinxxcontour} to write 

%
\begin{equation}
\begin{split}
\lim \limits_{R\to \infty} \lim \limits_{\epsilon\to 0} \int\limits_{[-\infty,\theta_1 -\epsilon ] \cup  [\theta_1+\epsilon ,\theta_2 -\epsilon]  \cup \dots [ \theta_n +\epsilon, \infty]}  \frac{dk}{2\pi}  \frac{ \exp\left( \frac{1}{2} i C(\vec s) k \right)}{i^n\prod_{\alpha=1}^n (k-\theta_\alpha)} =  0 - \lim \limits_{R\to \infty} \lim \limits_{\epsilon\to 0} \left( \int\limits_{\gamma_R}    + \sum_\beta  \int\limits_{\gamma_\beta} \right)\frac{dk}{2\pi}  \frac{ \exp\left( \frac{1}{2} i C(\vec s) k \right)}{i^n\prod_{\alpha=1}^n (k-\theta_\alpha)}
\end{split}
\end{equation}
The integral for the closed contour is 0, since there are no poles inside. Along the $\gamma_R$ large semi-circle, the integrand is bounded by $\mathcal{O}(R^{-n})$ for all $C(\vec s) \geq 0$ (even for $C(\vec s)=0$), therefore the integral on $\gamma_R$ vanishes as ${R \to \infty}$ for any $n\geq2$. Along a $\gamma_\beta$  semi-circle  we can parametrize $k$ as $k  = \epsilon e^{i \varphi} +\theta_\beta$, and thus $dk =   \epsilon e^{i \varphi} i d\varphi$, where $\varphi$ goes from $\pi$ to $0$. As a result we find 

\begin{equation}
\begin{split}
\lim \limits_{ \epsilon\to 0}\int\limits_{\gamma_\beta}\frac{dk}{2\pi}  \frac{ \exp\left( \frac{1}{2} i C(\vec s) k \right)}{i^N\prod\limits_{\alpha=1}^n (k-\theta_\alpha)}
 =  \lim \limits_{ \epsilon\to 0}\int\limits_{\pi}^0\frac{1}{2\pi}  \frac{ \epsilon e^{i \varphi} id \varphi}{\epsilon e^{i \varphi}} \frac{ \exp\left( \frac{1}{2} i C(\vec s) (\epsilon e^{i\varphi}+\theta_\beta) \right)}{i^n\prod\limits_{\alpha(\neq \beta) }^n (\epsilon e^{i \varphi} +\theta_\beta-\theta_\alpha)}= -\frac{1}{2i^{n-1}}  \frac{\exp( \frac{1}{2} iC(\vec s) \theta_\beta )}{\prod\limits_{\alpha(\neq \beta) }^n (\theta_\beta-\theta_\alpha)}.
\end{split}
\end{equation}
After simplifying we find that the integrand goes to a constant finite value as $\epsilon\to 0$, therefore the integral becomes trivial. Note that if $C(\vec{s})\leq 0$, a similar argument works by considering the contour shown in Figure~\ref{fig:contour2}.

In this case, for the $\gamma'_\beta$c contour   $k  = \epsilon e^{i \varphi} +\theta_\beta$,  $dk =   \epsilon e^{i \varphi} i d\varphi$ but $\varphi$  now goes from $\pi$ to $2\pi$. Therefore we have
\begin{equation}
\begin{split}
\lim \limits_{ \epsilon\to 0}\int\limits_{\gamma'_\beta}\frac{dk}{2\pi}  \frac{ \exp\left( \frac{1}{2} i C(\vec s) k \right)}{i^n\prod\limits_{\alpha=1}^n (k-\theta_\alpha)}
 =  \lim \limits_{ \epsilon\to 0}\int\limits_{\pi}^{2\pi}\frac{1}{2\pi}  \frac{ \epsilon e^{i \varphi} id \varphi}{\epsilon e^{i \varphi}} \frac{ \exp\left( \frac{1}{2} i C(\vec s) (\epsilon e^{i\varphi}+\theta_\beta) \right)}{i^n\prod\limits_{\alpha(\neq \beta) }^n (\epsilon e^{i \varphi} +\theta_\beta-\theta_\alpha)}= +\frac{1}{2i^{n-1}}  \frac{\exp( \frac{1}{2} iC(\vec s) \theta_\beta )}{\prod\limits_{\alpha(\neq \beta) }^n (\theta_\beta-\theta_\alpha)}.
\end{split}
\end{equation}
To verify that we get the same result using either contour for $C(\vec{s})=0$, we make use of the following identity
\begin{equation}
\begin{split}
 \sum_\beta \frac{1}{\prod_{\alpha (\neq\beta)} (\theta_\beta-\theta_\alpha)} \equiv 0. 
\end{split}
\label{eq:0identity}
\end{equation}
An elegant proof of this identity can be found in [\onlinecite{YANG2005102}]. Finally, we find


\begin{equation}
\begin{split}
z(\theta_1,\theta_2,\dots ,\theta_n)&=\sum_{\vec s \in \otimes^n\{-1,+1\}}  \exp\Big( -\frac{i}{2}  \sum_{\alpha=1}^n s_\alpha \theta_\alpha \Big)\Big(\prod\limits_\alpha s_\alpha\Big)  \text{sgn}(C(\vec s)) \frac{1}{2} \sum_\beta\frac{\exp( \frac{1}{2} iC(\vec s) \theta_\beta )}{i^{N-1}\prod\limits_{\alpha(\neq \beta) }^n (\theta_\beta-\theta_\alpha)},
\end{split}
\label{eq:genz1}
\end{equation}
where $\text{sgn}(C(\vec s))$ is the sign of $C(\vec s)$, with the added convention that $\text{sgn}(0)=0$. Rearranging the sums, this result can be rewritten as


\begin{equation}
\begin{split}
z(\theta_1,\theta_2,\dots ,\theta_n)&= \sum_\beta \sum_{\{s_1,\dots s_{\beta-1}, s_{\beta+1}\dots s_n\}} \frac{1}{2} \frac{ \exp\Big( \frac{i}{2}  \sum\limits_{\alpha(\neq \beta)} s_\alpha( \theta_\beta-\theta_\alpha) \Big)}{i^{n-1}\prod\limits_{\alpha(\neq \beta) }^n (\theta_\beta-\theta_\alpha) } \Big(\prod\limits_{\alpha(\neq \beta)} s_\alpha\Big)\sum_{s_\beta \in \{-1, +1\}}  s_\beta \, \text{sgn}(C(\vec s)).\end{split}
\label{eq:genz2}
\end{equation}
For fixed $\beta$ and fixed $ \{s_1,\dots s_{\beta-1}, s_{\beta+1}\dots s_N\}$, only $s_\beta \, \text{sgn}\big(C(\vec s)\big)$ depends on $s_\beta$. The two terms for $s_\beta=\pm1$ will cancel each other out unless $\text{sgn}(C(\vec{s}))$ changes sign when we change $s_\beta$. Thus, we only consider those configurations in the following. Our final expressions depend on the parity of $n$:


\subsection*{Case 1: $n$ odd}

For odd $n$, $C(\vec s)$ is also odd. For given $ \{s_1,\dots s_{\beta-1}, s_{\beta+1}\dots s_N\}$, $C(\vec s)$ will change sign upon changing $s_\beta$ only  if  $C(\vec s)= 1$ for $s_\beta=1$ and $ C(\vec s)=-1$ for $s_\beta=-1$, or equivalently when $\sum_{\alpha (\neq\beta)}s_\alpha=0$. For such terms  ${\prod\limits_{\alpha(\neq \beta)} =(-1)^{(n-1)/2}}$, and  $\sum\limits_{s_\beta} s_\beta \, \text{sgn}\big(C(\vec s)\big)=2$, therefore we end up with
\begin{equation}
\begin{split}
z(\theta_1,\theta_2,\dots ,\theta_n)&= \sum_\beta \sum_{\substack{\{s_1,\dots s_{\beta-1}, s_{\beta+1}\dots s_n\} \\ \sum\limits_{\alpha(\neq \beta)}s_\alpha=0}} \frac{ \exp\Big(- \frac{i}{2}  \sum\limits_{\alpha(\neq\beta)} s_\alpha \theta_\alpha \Big)}{\prod\limits_{\alpha(\neq \beta) }^n (\theta_\beta-\theta_\alpha) }. \end{split}
\label{eq:oddgenz1}
\end{equation}
 Since $z(\vec \theta_1, \theta_2,\dots \theta_n)$ should be real, we can just take the real part of Eq.~\eqref{eq:oddgenz1}, or equivalently we can combine the terms of $ \{s_1,\dots s_{\beta-1}, s_{\beta+1}\dots s_n\}$ and $ \{-s_1,\dots -s_{\beta-1}, -s_{\beta+1}\dots -s_n\}$ to arrive at

\begin{equation}
\begin{split}
z(\theta_1,\theta_2,\dots ,\theta_n)&= \sum_\beta \sum_{\substack{\{s_1,\dots s_{\beta-1}, s_{\beta+1}\dots s_n\} \\ \sum\limits_{\alpha(\neq \beta)}s_\alpha=0}} \frac{ \cos\Big(\frac{1}{2}  \sum\limits_{\alpha(\neq\beta)} s_\alpha \theta_\alpha \Big)}{\prod\limits_{\alpha(\neq \beta) }^n (\theta_\beta-\theta_\alpha) }. \end{split}
\label{eq:oddgenz2}
\end{equation}
This agrees with (\ref{st:result1}).

\subsection*{Case 2: $n$ even}

For even $n$, $C(\vec s)$ is also even. In this case we only need to consider the terms where $C(\vec s) = 2 $ for $s_\beta=1$ and becomes $C(\vec s)=0$ for $s_\beta=-1$, or similarly when  $C(\vec s) = - 2  $ for $s_\beta=-1$ and becomes $C(\vec s)=0$ for $s_\beta=1$. The former corresponds to cases when $\sum_{\alpha(\neq \beta)}s_\alpha =1$ and $\prod_{\alpha(\neq \beta)} s_\alpha= (-1)^{n/2-1}$ , while the latter is when  $\sum_{\alpha(\neq \beta)} s_\alpha=-1$ and  $\prod_{\alpha(\neq \beta)} s_\alpha= (-1)^{n/2}$. In both cases the $\sum_\beta s_\beta\, \text{sgn}(C(\vec s))$ term gives 1:

\begin{equation}
\begin{split}
z(\theta_1,\theta_2,\dots ,\theta_n)=&{} \sum_\beta \sum_{\substack{\{s_1,\dots s_{\beta-1}, s_{\beta+1}\dots s_n\}\\ \sum\limits_{\alpha(\neq \beta)}s_\alpha=1}} \frac{1}{2i} \frac{ \exp\Big( \frac{i}{2}  \sum\limits_{\alpha(\neq \beta)} s_\alpha( \theta_\beta-\theta_\alpha) \Big)}{\prod\limits_{\alpha(\neq \beta) }^n (\theta_\beta-\theta_\alpha) } \\
&+ \sum_\beta \sum_{\substack{\{s_1,\dots s_{\beta-1}, s_{\beta+1}\dots s_n\}\\ \sum\limits_{\alpha(\neq \beta)}s_\alpha=-1}} \frac{i}{2} \frac{ \exp\Big( \frac{i}{2}  \sum\limits_{\alpha(\neq \beta)} s_\alpha( \theta_\beta-\theta_\alpha) \Big)}{\prod\limits_{\alpha(\neq \beta) }^n (\theta_\beta-\theta_\alpha) }.
\end{split}
\label{eq:evengenz1}
\end{equation}
For a configuration $\{s_1,\dots s_{\beta-1}, s_{\beta+1}\dots s_n\}$, with $\sum\limits_{\alpha(\neq \beta)}s_\alpha =\pm 1$, we can uniquely determine an $s_\beta =\mp 1$ for which $\sum_\alpha s_\alpha=0$:

\begin{equation}
\begin{split}
z(\theta_1,\theta_2,\dots ,\theta_n)=&\mathbin{\phantom{+}} \sum_\beta \sum_{\substack{\{s_1,\dots s_{\beta-1},s_\beta, s_{\beta+1}\dots s_n\}\\ \sum\limits_{\alpha}s_\alpha =0, s_\beta=-1}} \frac{1}{2i} \frac{ \exp\Big( \frac{i}{2}  \sum\limits_{\alpha} s_\alpha( \theta_\beta-\theta_\alpha) \Big)}{\prod\limits_{\alpha(\neq \beta) }^n (\theta_\beta-\theta_\alpha) } \\
&+ \sum_\beta \sum_{\substack{\{s_1,\dots s_{\beta-1},s_\beta, s_{\beta+1}\dots s_n\}\\ \sum\limits_{\alpha}s_\alpha=0, s_\beta=1}} \frac{i}{2} \frac{ \exp\Big( \frac{i}{2}  \sum\limits_{\alpha} s_\alpha( \theta_\beta-\theta_\alpha) \Big)}{\prod\limits_{\alpha(\neq \beta) }^n (\theta_\beta-\theta_\alpha) }\\
 \end{split}
 \label{eq:evengenz2}
\end{equation}
\begin{equation}
\begin{split}
=&\mathbin{\phantom{+}}  \sum_{\substack{\{\vec s\}\\ \sum\limits_{\alpha}s_\alpha =0}} \left[   \sum_{\substack{\beta:\\s_\beta=-1}}  \frac{1}{2i} \frac{ \exp\Big( -\frac{i}{2}  \sum\limits_{\alpha} s_\alpha \theta_\alpha \Big)}{\prod\limits_{\alpha(\neq \beta) }^n (\theta_\beta-\theta_\alpha) } - \sum_{\substack{\beta:\\s_\beta=1}} \frac{1}{2i} \frac{ \exp\Big( -\frac{i}{2}  \sum\limits_{\alpha} s_\alpha \theta_\alpha \Big)}{\prod\limits_{\alpha(\neq \beta) }^n (\theta_\beta-\theta_\alpha) }\right]
 \end{split}
 \label{eq:evengenz3}
\end{equation}
Once again we can argue that $z(\theta_1, \dots \theta_n)$ has to be real, so we can just take the real part of the above. Or we arrive to the same result by combining the $\vec s$ and $-\vec s$ terms,

\begin{equation}
\begin{split}
z(\theta_1,\theta_2,\dots ,\theta_n)=&\mathbin{\phantom{+}}  \sum_{\substack{\{\vec s\}\\ \sum\limits_{\alpha}s_\alpha =0}} \frac{1}{2}\sin\Big(\frac{1}{2}  \sum\limits_{\alpha} s_\alpha \theta_\alpha \Big)  \left[   \sum_{\substack{\beta:\\s_\beta=1}}   \frac{1}{\prod\limits_{\alpha(\neq \beta) }^n (\theta_\beta-\theta_\alpha) } - \sum_{\substack{\beta:\\s_\beta=-1}}  \frac{1}{\prod\limits_{\alpha(\neq \beta) }^n (\theta_\beta-\theta_\alpha) }\right].
\end{split}
\label{eq:evengenzv4}
\end{equation}
Making use of the identity  in Eq.~\eqref{eq:0identity}, we end up with
\begin{equation}
\begin{split}
z(\theta_1,\theta_2,\dots ,\theta_n)=&\mathbin{\phantom{+}}  \sum_{\substack{\{\vec s\}\\ \sum\limits_{\alpha}s_\alpha =0}} \sin\Big(\frac{1}{2}  \sum\limits_{\alpha} s_\alpha \theta_\alpha \Big)  \left[   \sum_{\substack{\beta:\\s_\beta=1}}   \frac{1}{\prod\limits_{\alpha(\neq \beta) }^n (\theta_\beta-\theta_\alpha)} \right],
\end{split}
\label{eq:evengenzv5}
\end{equation}
which proves (\ref{st:result2}).

\bibliographystyle{apsrev4-1}
\bibliography{sunbib.bib}

\end{document}